\documentclass{aa}  

\usepackage{graphicx}
\usepackage{txfonts}
\usepackage[breaklinks, colorlinks, citecolor=blue]{hyperref}
\usepackage{xcolor}

\usepackage{xpatch}
\usepackage{hyperref}
\hypersetup{
	colorlinks=true,
	breaklinks=true,
	citecolor=blue,
	allcolors=blue,
	frenchlinks=true
}

\makeatletter
\xpatchcmd\NAT@citex
{%
	\@citea\NAT@hyper@{%
		\NAT@nmfmt{\NAT@nm}%
		\hyper@natlinkbreak{\NAT@aysep\NAT@spacechar}{\@citeb\@extra@b@citeb}%
		\NAT@date
	}%
}
{%
	\@citea
	\NAT@nmfmt{\NAT@nm}%
	\NAT@aysep\NAT@spacechar
	\NAT@hyper@{\NAT@date}%
}
{}{}
\xpatchcmd\NAT@citex
{%
	\@citea\NAT@hyper@{%
		\NAT@nmfmt{\NAT@nm}%
		\hyper@natlinkbreak{\NAT@spacechar\NAT@@open\if*#1*\else#1\NAT@spacechar\fi}%
		{\@citeb\@extra@b@citeb}%
		\NAT@date
	}%
}
{
	\@citea
	\NAT@nmfmt{\NAT@nm}%
	\NAT@spacechar\NAT@@open\if*#1*\else#1\NAT@spacechar\fi
	\NAT@hyper@{\NAT@date}%
}
{}{}
\makeatother

\makeatletter
\renewcommand*\aa@pageof{, page \thepage{} of \pageref*{LastPage}}
\makeatother

\newcommand{\bs}[1]{\boldsymbol{#1}}
\newcommand{\bcdot}{\boldsymbol{\cdot}}
\newcommand{\bnabla}{\boldsymbol{\nabla}}

\begin{document} 

   \title{Effective cosmic ray diffusion in multiphase galactic environments}
   \titlerunning{Effective cosmic ray diffusion in multiphase galactic environments}
   \authorrunning{Thomas et al.}

   \author{T. Thomas\inst{1}\fnmsep\thanks{\email{tthomas@aip.de}},
          C. Pfrommer\inst{1},
          R. Pakmor\inst{2},
          R. Lemmerz\inst{3,1},
          and
          M. Shalaby\inst{4,5,6,1}
            \fnmsep\thanks{{Horizon AstroPhysics Initiative (HAPI) Fellow}}
          }
   \institute{
   Leibniz Institute for Astrophysics Potsdam (AIP), An der Sternwarte 16, 14482 Potsdam, Germany
    \and
    Max Planck Institute for Astrophysics, Karl-Schwarzschild-Strasse 1, 85740 Garching, Germany
    \and
    Department of Physics, University of Wisconsin-Madison, Wisconsin 53706, USA
    \and
Waterloo Centre for Astrophysics, University of Waterloo, Waterloo, ON N2L 3G1, Canada    \and
Perimeter Institute for Theoretical Physics, 31 Caroline St.~N., Waterloo, Ontario N2L 2Y5, Canada    \and
Department of Physics and Astronomy, University of Waterloo, Waterloo, ON, N2L 3G1, Canada}

   \date{Received ...; accepted ...}

% \abstract{}{}{}{}{} 
% 5 {} token are mandatory
 
  \abstract
  % context heading (optional)
  % {} leave it empty if necessary  
   { 
   Cosmic-ray (CR) feedback is widely recognized as a key regulator of galaxy formation. After being accelerated at supernova remnant shocks, CRs propagate through the interstellar medium (ISM), establishing smooth large-scale distributions and driving galactic outflows. The efficiency of this feedback is controlled by the effective transport speed of the CR population, which in turn depends on the competition between CR-driven plasma instabilities and wave damping processes that vary strongly with ISM phase. In cold, dense gas, ion-neutral damping dominates, whereas in warm, diffuse environments, weaker non-linear Landau damping prevails, leading to enhanced CR scattering and slower transport. To investigate these effects, we employ the moving-mesh code \textsc{Arepo} and model CR transport using a two-moment description within the multiphase ISM framework \textsc{Crisp}, which self-consistently computes CR diffusion coefficients and transport velocities from coarse-grained plasma physics. The intrinsic CR diffusion coefficient depends inversely on the scattering rate of CRs and Alfv\'en waves, covering 15 orders of magnitude. In contrast, we show that the effective CR diffusion coefficient, which quantifies the propagation speed of CRs through the ISM, converges toward the canonical range of $10^{28}$–$10^{29}~\mathrm{cm}^2~\mathrm{s}^{-1}$. Simulations with only non-linear Landau damping yield transport rates up to an order of magnitude slower than those including both Landau and ion-neutral damping. Overall, CR transport speeds increase systematically with gas density, for which we provide a density-dependent fit of the effective CR diffusion coefficient. We demonstrate that, despite strong ion-neutral damping in the cold and warm phases of the galactic disk, CRs rapidly smooth out their gradient and are transported at speeds only a few times the local Alfv\'en speed as they traverse alternating ISM phases on their way out of the galaxy.  This approach enables us to investigate how self-consistent CR transport impacts CR induced non-thermal observables in the future.
 }


   \maketitle

%--------------------------------------------------------------------------------------------------------------------
%--------------------------------------------------------------------------------------------------------------------

\section{Introduction}

In the cosmological standard model, galaxies assemble within dark matter halos as baryonic gas cools and condenses to form stars. Yet, despite the short radiative cooling timescales, only a small fraction of this gas is converted into stars -- less than 1\% in dwarf galaxies \citep{2010Moster}. This inefficiency points to the critical role of stellar feedback in regulating star formation and preventing excessive gas collapse \citep{2015Somerville,2017Naab}. Such feedback encompasses a range of processes, including mechanical energy injection from supernova (SN) explosions, photoionization and radiation pressure from massive stars, and the influence of CRs accelerated at supernova shock fronts.

In recent years, specifically CR feedback has emerged as a key ingredient in shaping galaxy evolution \citep{2023Ruszkowski}. As CRs propagate away from their acceleration sites, they establish a relatively smooth distribution throughout the ISM. Their outward diffusion from the galactic disk generates a pressure gradient in the halo, capable of driving large-scale galactic winds \citep{2014Salem} that lift interstellar gas out of the gravitational potential well and enrich the circumgalactic medium (CGM). Interestingly, CRs are able to move the hot and cold phases of the ISM into the halo via their pressure gradient and magnetic bottlenecks \citep{2017Wiener,2021Thomas}, respectively. In addition, the high specific energy of CRs enables slow cooling, allowing them to dominate the inner CGM pressure and thereby regulate angular momentum transport and disk sizes \citep{2020Buck,2020HopkinsII,2025Bieri}. The mechanism of CR-driven winds has been successfully demonstrated with simplified one-dimensional settings \citep{1975Ipavich,1991Breitschwerdt,1993Breitschwerdt,2016Recchia,2022QuataertII} as well as across a broad range of numerical simulations, including high-resolution ISM tall-box simulations \citep{2016Girichidis,2018Girichidis,Simpson2016,Simpson2023,2018Farber,2022Armillotta,2024Armillotta,2025Sike}, isolated disk galaxy models \citep{2013Hanasz,2013Booth,2014Salem,2017Ruszkowski,2018Butsky,2020Dashyan,2022Farcy,2023Thomas,2025Thomas,2025Kjellgren}, galaxy formation from collapsing gas clouds embedded in dark matter halos \citep{2012Uhlig,2016PakmorIII,2017Wiener,2018Jacob,2024Girichidis}, and fully cosmological galaxy formation simulations \citep{2016Salem,2020Buck,2020HopkinsII,2025Bieri}.

Owing to their quasi-collisionless nature, CRs exhibit distinct transport behaviour in the various ISM phases they permeate. Magnetic fields play a central role in mediating both CR dynamics and transport. GeV-energy CR protons dominate the CR energy budget and interact with magnetic fields in several key ways: (i) CRs primarily propagate along large-scale magnetic field lines \citep{2013Zweibel,2017Zweibel,2021Hanasz}, as their gyroradii are much smaller than typical magnetic correlation lengths on galactic and ISM scales \citep{2022Pfrommer}. Because magnetic fields are effectively frozen into the plasma, bulk gas motions advect CRs along with the field lines. (ii) CRs scatter off magnetic turbulence through resonant and non-resonant interactions, a process that macroscopically manifests as diffusion \citep{2004Yan,2011Yan}. In certain conditions, CRs can become trapped in localized regions by magnetic mirrors or microinstabilities, causing their transport to deviate from classical diffusion and follow a Lévy-flight-like process \citep{2021Lazarian,2025Ewart}. (iii) CRs can also self-excite small-scale magnetic fluctuations through plasma instabilities and subsequently scatter off the waves they generate \citep{1969Kulsrud,2019Bai,2021Shalaby,2023Shalaby,2025Lemmerz}. This process results in nearly pure streaming when CRs scatter frequently, whereas (somewhat) less frequent scattering leads to a combination of streaming and diffusion, provided the CRs remain well coupled to the magnetic field \citep{2020Thomas}.

Modelling CR transport has a long history, with early approaches relying on phenomenological, energy-dependent isotropic diffusion in static Galactic models to reproduce many features of the observed CR composition, energy spectra, and gamma-ray emission \citep{1998Moskalenko,2002Moskalenko,2000Strong,2007Strong,2008Evoli,2019Evoli}. However, the improved level of detail provided by Fermi-LAT data has revealed several tensions with these models. Chief among them is the so-called “gradient problem,” which refers to the difficulty of explaining the centrally peaked gamma-ray emission when the CR source distribution peaks at Galactocentric radii of  $\approx$3~kpc \citep{2002Breitschwerdt,2011Ackermann,2012Evoli}. Moreover, such phenomenological models are in conflict with plasma-kinetic theory of CR transport in galaxies, which predicts anisotropic, magnetically guided propagation and a non-negligible role for streaming at GeV energies \citep{2022Kempski}. Indeed, a combination of CR streaming and diffusion appears necessary to reproduce the observed radio brightness profiles of the non-thermal ``radio harp'' filaments \citep{2020Thomas} located in the Central Molecular Zone of the Milky Way \citep{2022Heywood}. More recently, CR transport has been modelled directly in CR–magnetohydrodynamical (CRMHD) simulations of galaxy evolution to connect CR physics with multi-wavelength observables \citep{2017bPfrommer,2021WerhahnIII,2021WerhahnII,2021WerhahnI,2023Werhahn,2021Ogrodnik,2022Hopkins_spectra,2024Chiu}. In particular, the CR composition, energy spectra, and
gamma-ray emission in these simulations have highlighted significant challenges, with several predicted observables still in tension with observational data \citep{2022Hopkins_transport_problems}, underscoring the need for improved CR transport prescriptions.

Post-processing high-resolution simulations of the multiphase ISM with detailed CR transport modelling, \citet{2025Armillotta} demonstrated that including both advection and momentum-dependent diffusion -- where the scattering coefficients are computed locally from the steady-state balance between streaming-driven Alfvén wave growth and damping processes -- yields CR spectra in excellent agreement with observations. This motivates our present study of CR transport in the multiphase ISM of an entire galaxy. To this end, we employ the multiphase ISM framework \textsc{Crisp} \citep{2025Thomas}, which self-consistently tracks the non-equilibrium chemistry of 12 species, includes CR and photoelectric dust heating, models isotropically distributed supernovae and stellar winds, and follows radiative transfer coupled to metal and dust evolution. Importantly, \textsc{Crisp} integrates a two-moment CRMHD framework that captures key plasma-kinetic effects on macroscopic scales by evolving both the CR energy and momentum densities, as well as the energy density of small-scale resonant Alfvén waves propagating in both directions along the mean magnetic field. These waves grow via the CR streaming instability and are damped by multiple microphysical processes, thereby regulating the local CR scattering rate. With this approach, we aim to investigate how self-consistent ISM dynamics in a global galaxy impacts CR transport in the different ISM phases. 

In contrast to our previous works \citep{2023Thomas}, we include the effect on ion-neutral damping on CR transport in the present study. The rate at which ion-neutral damping is impacting CR transport depends heavily on the thermodynamical state of the ISM which is simulated by the \textsc{Crisp} framework. To quantify its impact, we compare simulations that include the ion-neutral damping process to those that do not follow this process in terms of quantities that statistically describe CR transport such as the streaming speed and the diffusion coefficient. This paper is structured as follows: we provided an overview of the relevant CR-transport terminology and physics in Sec.~\ref{sec:introduction_physics} and describe the difference between the intrinsic and effective diffusion coefficient in Sec.~\ref{sec:intrinsic_vs_effective}. We present our simulation setup in Sec.~\ref{sec:simulation_setup}, give a broad overview of the resulting galaxies in Sec.~\ref{sec:overview}, and start to analyse the effective CR diffusion coefficient in Sec.~\ref{sec:effective CR transport}. We then focus on the statistics of this quantity in Sec.~\ref{sec:intrinsic_vs_effectiv_near_disk} and put our findings into the context of the contemporary literature in Sec.~\ref{sec:discussion}. We conclude and summarize our results in Sec.~\ref{sec:conclusion}.

\section{Simulated CR physics}
\label{sec:introduction_physics}

In this section, we introduce and discuss key aspects of CR transport physics, focusing on what will be relevant for our analysis of CR propagation within the galactic disk and galactic outflows.

\subsection{CRs and Alfv\'en waves}

We focus on the collective behaviour of GeV CR protons and restrict ourselves to the canonical self-confinement picture of CR transport. Therein, the CR transport is regulated by the electro-magnetic interaction between CRs and small-scale Alfv\'en waves which alters the CR's pitch angle (the angle between the CR momentum and the magnetic field), effectively scattering them off electromagnetic plasma waves. These scattering processes cause near isotropization of the CR momentum vectors in the Alfv\'en wave frame and, in consequence, slows down the mean CR velocity of the intrinsically relativistic particles to almost the local Alfv\'en speed. In this picture, these Alfv\'en waves are provided by the CRs themselves and are excited through the gyroresonant CR streaming instability \citep{1969Kulsrud,2023Shalaby}. This CR-excitation of Alfv\'en waves and subsequent scattering of CRs creates a self-regulating feedback loop which determines how fast CRs are transported throughout the galaxy and its galactic wind.

To discuss the main CR-transport processes, we condense the complexity of the CRMHD equation system \citep[the full set given by][]{2019Thomas} down to 
\begin{align}
    \frac{\partial \varepsilon_\mathrm{cr}}{\partial t} + \bnabla \bcdot (f_\mathrm{cr} \bs{b}) &= -\varv_\mathrm{a,i} \frac{\nu}{c^2} \left[f_\mathrm{cr} - \varv_\mathrm{a,i} (\varepsilon_\mathrm{cr} + P_\mathrm{cr}) \right], \\
    \frac{1}{c^2} \frac{\partial f_\mathrm{cr}}{\partial t} + (\bs{b} \cdot \bnabla) P_\mathrm{cr} &= -\phantom{\varv_\mathrm{a,i}} \frac{\nu}{c^2} \left[f_\mathrm{cr} - \varv_\mathrm{a,i} (\varepsilon_\mathrm{cr} + P_\mathrm{cr}) \right] \label{eq:cr_flux} \\
    \frac{\partial \varepsilon_\mathrm{a}}{\partial t}   &= +\varv_\mathrm{a,i} \frac{\nu}{c^2} \left[f_\mathrm{cr} - \varv_\mathrm{a,i}(\varepsilon_\mathrm{cr} + P_\mathrm{cr}) \right] - \Gamma \varepsilon_\mathrm{a} \label{eq:ealf},
\end{align}
where the various quantities have the following meaning:
\begin{align}
     \varepsilon_\mathrm{cr} \, \& \, P_\mathrm{cr} &\quad-\quad \text{cosmic ray energy density \& pressure,} \nonumber \\
     f_\mathrm{cr} &\quad-\quad \text{cosmic ray energy flux density,}  \nonumber \\
     \varepsilon_\mathrm{a} &\quad-\quad \text{Alfv\'en wave energy density,} \nonumber \\
     \varv_\mathrm{a,i} &\quad-\quad \text{ion Alfv\'en (wave) speed,}  \nonumber \\
     c &\quad-\quad \text{speed of light,} \nonumber \\
     \bs{b} = \bs{B} / B &\quad-\quad \text{direction of the magnetic field } \bs{B}, \nonumber \\
     \nu &\quad-\quad \text{scattering frequency of CRs by Alfv\'en waves}, \nonumber  
\end{align}
and the damping rate of Alfv\'en wave energy density is given by
\begin{equation}
    \Gamma = \Gamma_\mathrm{NLLD} + \Gamma_\mathrm{IND},
\end{equation}
which is the sum of damping rates provided by non-linear Landau damping ($\Gamma_\mathrm{NLLD}$) and ion-neutral damping ($\Gamma_\mathrm{IND}$). In addition to these two damping processes, others can also influence the amount of energy stored in small-scale Alfv\'en waves \citep{2013Zweibel}. We focus on those two processes because other damping mechanisms are less powerful or too uncertain to be numerically implemented. To study the relative impact of ion-neutral damping, we consider a simulation model that includes this effect and compare it to a model that only incorporates non-linear Landau damping. The latter process is a prerequisite for successfully employing the two-moment scheme of \citet{2021Thomas}, owing to its non-linear character ($\Gamma_\mathrm{NLLD} \propto \varepsilon_\mathrm{a}$). This ensures that, even in cases of rapid growth of small-scale Alfvén waves, the energy density stored in these waves ($\varepsilon_\mathrm{a}$) cannot exceed a physically imposed upper limit.

This equation set is far from complete and neglects how CRs interact with the thermal gas, the MHD equations, various inertia terms, the presence of Alfv\'en waves that travel and scatter in both directions of the magnetic field -- but it is sufficient to explore and explain CR transport along magnetic field lines.

The left-hand sides of these equations tells us that the CR energy flux dictates how fast CR energy is transported along magnetic field lines, while the CR energy flux is generally seeded by CR pressure gradients along the magnetic field. The latter term seeds the general motion of CR energy from over-pressurized to under-pressurized regions. 

The workings of the source terms on the right-hand side is best understood when we identify the equilibria of the equations set. Ignoring the time derivative in Eq.~\eqref{eq:cr_flux} and rearranging the result to yield an expression for $f_\mathrm{cr}$ gives 
\begin{align}
    f_\mathrm{cr} = \varv_\mathrm{a,i} (\varepsilon_\mathrm{cr} + P_\mathrm{cr}) - \kappa_\mathrm{cr} (\bs{b} \bcdot \bnabla) \varepsilon_\mathrm{cr},
\end{align}
where we introduced the \textbf{intrinsic CR diffusion coefficient} $\kappa_\mathrm{cr} = c^2 / \nu (\gamma_\mathrm{cr} - 1)$ with $\gamma_\mathrm{cr}$ being the adiabatic index of the CR fluid. The first term tells us that CRs are transported preferentially with the Alfv\'en speed along magnetic field lines. This effect is referred to as \textbf{CR streaming}. The second term adds a contribution similar to Fick's first law for Brownian diffusion which leads to \textbf{CR diffusion} along magnetic field lines. The interpretation of the underlying physical processes that gives rise to both Brownian and CR diffusion are similar but not equal. In the context of Brownian motion, the microscopic reason for diffusion are frequent collisions of heavy particles with abundant lighter particles that build up a fluid surrounding the heavier particle. The (assumed-to-be) stochastic nature of these collisions leads to a net flux of particles from regions of high to low concentration of heavy particles. The frequency of such collisions is mostly determined by the temperature or in other words the thermal energy content of the fluid composed of lighter particles. CRs are not embedded in a sea of lighter particles, but are subject to collisions with the electro-magnetic fields of Alfv\'en waves, which leads to CR-wave scattering mediated by the Lorentz force. Similarly, the scattering rate $\nu$ is determined by the energy content of the scattering medium, i.e.\ the Alfv\'en waves in the case of CRs, so that we obtain 
\begin{equation}
    \nu = \frac{3 \pi}{16} \Omega \frac{\varepsilon_\mathrm{a}}{\varepsilon_\mathrm{B}} \propto \varepsilon_\mathrm{a},
\end{equation}
where $\Omega$ is the gyrofrequency of the CRs and $\varepsilon_\mathrm{B} = B^2 / 8\pi$ is the energy density of the magnetic field. This implies that more scatterings will take place in regions where there are many Alfv\'en waves, implying that the diffusive particle flux is slowed down. It is thus paramount to have an accurate estimate for $\varepsilon_\mathrm{a}$ in order to know how fast CR diffusion is. A candidate estimate can be derived when we set the right-hand side of Eq.~\eqref{eq:ealf} to zero. In practice, this condition is realized when the growth of Alfv\'en waves by CR streaming is precisely counteracted by wave damping processes. This can only be achieved if the first term on the right-hand side of Eq.~\eqref{eq:ealf} (that is proportional to $f_\mathrm{cr}$) is positive as otherwise no equilibrium arises, i.e, a net cancellation of all processes affecting the time evolution of wave energy. A scenario where this is \textit{not} the case are Alfv\'en-wave dark regions as noted by \citet{2023Thomas}, where CRs stream slower than the wave propagation speed and thus damp the waves themselves upon interacting with them. We will encounter such regions in our analysis of CR transport.

\subsection{Ion-neutral damping}

Ion-neutral damping is best understood by 1) fully embracing the frozen-in approximation of MHD and assuming that the magnetic fields are directly tied to the thermal ions while following their motions, and 2) considering a neutral atom at rest and a moving ion that approaches the neutral. Even though the ion and the neutral atom do not directly interact via electromagnetic forces, the approaching ion ever so slightly displaces the bound electrons on the neutral from its positively charged core. This displacement induces a dipole electric field that can back-react on the approaching ion. The dipole electric field, and hence, the neutral now exerts a force on the ion and transfers some of the ion's momentum to the neutral atom. By altering the ion's motion, this interaction also affects the motion of the frozen-in magnetic field and thus is able to, for example, damp Alfvén waves that would otherwise scatter CRs. 

The ISM is a multi-component gas consisting of multiple primordial and metallic ions and neutrals. The total ion-neutral damping rate for such a mixture is given by:
\begin{align}
    \Gamma_\mathrm{IND} = \frac{\sum_{i,n} n_i n_n \langle \sigma \varv \rangle_\mathrm{mt}}{\sum_i n_i}, 
\end{align}
where the sums range over ions $i$ and neutral particles $n$, $n_i$ is the number density of the ions, $n_n$ is the number density of the neutral particles, and $\langle \sigma \varv \rangle_\mathrm{mt}$ is the average momentum transfer cross-section of their interaction in units of cm$^{3}$ s$^{-1}$. We account for 16 possible collisions, namely 
\begin{itemize}
    \item \ion{H}{I} $\longleftrightarrow$ H$_2$, \ion{H}{I}, and \ion{He}{I}
    \item \ion{H}{II} $\longleftrightarrow$ \ion{C}{I}, \ion{O}{I}, and \ion{Si}{I}
    \item \ion{H}{I} and H$_2$ $\longleftrightarrow$ \ion{He}{II}, and \ion{He}{III}    
    \item \ion{H}{I} and H$_2$ $\longleftrightarrow$ \ion{C}{II}, \ion{O}{II}, and \ion{Si}{II}
\end{itemize}
and thus those ``first-order'' interactions where one of the collision partners is either a hydrogen atom or ion. We source the momentum transfer cross-sections from \citet{2008Pinto}, using their fits where available or using Langevin's approximation with atomic polarizabilities taken from \citet{1991Miller}.

\subsection{Focus on comoving transport}

Much like in the preceding discussion, we will be neglecting the effects of advective CR transport and solely focus on the properties of CR transport along magnetic field lines in the remainder of this paper. Nevertheless, advective CR transport is fully included in our simulations and is a principal CR transport mechanism. Analysis of the relative importance of advective versus other transport modes has been conducted for tall-box ISM simulations \citep{2021Armillotta} and global galaxy simulations \citep{2023Thomas, 2024Girichidis}.

\section{Intrinsic and effective diffusion}
\label{sec:intrinsic_vs_effective}

In our discussion of CR streaming and diffusion in Sec.~\ref{sec:introduction_physics}, we treated both processes as separate entities influencing CR transport. Their key defining quantities are the streaming speed and the diffusion coefficient, which both have different units and physical interpretations, which makes it difficult to directly compare both.  To obtain a statistical understanding of how fast the CR energy is transported through the ISM and the inner CGM, we use the \textbf{effective CR diffusion coefficient} $\kappa_\mathrm{eff}$, as defined via:
\begin{equation}
    \kappa_\mathrm{eff} = - \frac{f_\mathrm{cr}}{\bs{b} \bcdot \bnabla \varepsilon_\mathrm{cr}},
    \label{eq:kappa_eff}
\end{equation}
which is independent of the streaming speed and intrinsic diffusion coefficient, but characterizes the realized CR transport with a single quantity that is unconnected to the streaming or diffusion picture. Nevertheless, this quantity has the useful property that, in the absence of additional transport processes beyond diffusion, the intrinsic and effective diffusion coefficients are identical. Equivalently, we could define an effective CR transport speed, which contains the same amount of information and coincides with the streaming speed in the absence of diffusion. The effective CR diffusion coefficient is more useful, since its parameterization can be directly applied in numerical simulations that model CR transport via diffusion.

The discussion so far has assumed that diffusion actively contributes to CR transport. However, this might not be the case whenever CRs are scattered infrequently or not at all. In such situations, the Brownian motion picture of CR transport is not an appropriate description and the notion of intrinsic diffusion breaks down. Because $\kappa_\mathrm{cr} \propto \nu^{-1}$, the intrinsic diffusion coefficient becomes large for infrequent scatterings, implying that CR diffusion and thereby CR transport should be faster. In this regime, the diffusion approximation is no longer valid and CRs instead move ballistically, with particles following individual trajectories and collective behaviour is absent. A clear inconsistency emerges in the limit of vanishing scattering: the diffusion coefficient formally diverges, implying infinitely fast CR transport that would surpass the speed of light, which is unphysical. Vanishing CR scattering rates can be found in Alfv\'en-wave dark regions \citep{2023Thomas}, hence, demonstrating that this edge case is not an abstract academic exercise. The remedy lies in recognizing that the diffusion description no longer reflects physical reality and thus ceases to be applicable.

For the Alfv\'en-wave dark regions, the mentioned contradiction becomes even worse: while the intrinsic diffusion coefficient is high (in our simulations, it reaches a numerical floor value in excess of $10^{40}~\mathrm{cm}^2~\mathrm{s}^{-1}$) the mean transport velocity of the CR population is sub-Alfv\'enic and hence slow \citep{2023Thomas}. While ballistic behaviour is not described by any fluid theory, the two-moment theory at least can capture the absence of CR diffusion and the slow collective CR transport. As the intrinsic CR diffusion coefficient loses this physical relevance, it is automatically not affecting the evolution of CR transport any more.

Even in this extreme, yet physically realized regime, the effective CR diffusion coefficient $\kappa_\mathrm{eff}$ can still be defined. It effectively answers the question: ``What diffusion coefficient would reproduce the CR transport currently realized?'' In this sense, $\kappa_\mathrm{eff}$ encodes the instantaneous (possibly sub-Alfvénic) CR transport speed and reduces the complexity of the two-moment description to an equivalent one-moment diffusion equation.

\section{Simulation setup}
\label{sec:simulation_setup}

The simulations in this study were performed with the moving-mesh code \textsc{Arepo} \citep{2010Springel,2016PakmorII}, including its MHD module \citep{2011Pakmor,2013Pakmor} and CR module \citep{2017Pfrommer}, along with its extension for two-moment CRMHD \citep[][]{2019Thomas, 2021Thomas, 2022Thomas}. The galaxy's evolution is modelled within the \textsc{Crisp} framework, for which this section provides a brief overview. The NLLD+IND galaxy has already been presented in \citet{2025Thomas} and \citet{2024Chiu}. 
A detailed account of all used physical and numerical models will be given in a forthcoming paper by Thomas et al. (in prep.).

Our initial conditions consist of a gaseous and a stellar galactic disk, both with a radial scale length of 5 kpc and a vertical scale height of 0.5 kpc. The gaseous disks have a total mass of $0.8 \times 10^{10}~\mathrm{M}_\odot$ while the stellar disks weigh $3.2 \times 10^{10}~\mathrm{M}_\odot$. Both disks are sampled with discrete particles, each with an initial mass of 1000 M$_\odot$. This is also the target mass for \textsc{Arepo}'s refinement and derefinement mechanism that keeps the masses of gas particles around a predefined value \citep{2010Springel}. On top of this, we also employ a super-Lagrangian scheme to boost the numerical resolution inside the inner-CGM and use the same resolution targets as \citet{2025Thomas}. The galactic disk was embedded inside a Hernquist halo \citep{1990Hernquist} with mass $10^{12}~\mathrm{M}_\odot$, concentration $C = 7$ and baryon mass fraction $f_b = 0.155$ which we assumed to be static.   

We did not attempt to model a consistent CGM but embedded the galactic disks inside a hot and low-density ($n_\mathrm{H} = 10^{-6}~\mathrm{cm}^{-3}$) background medium. The ensuing galactic winds from the simulated galaxies freely run into the quasi-vacuum and do not need to work against the static pressure of the CGM or infalling material as realistic galactic winds would do. The domain of applicability of our simulations is thus restricted to the vicinity of the galactic disks, where the dynamics are determined by the galactic wind. We will not analyse our simulations above galactic heights $z \gtrsim 100$ kpc where this is certainly not the case. 

The magnetic field was initialized with two components. The first component was a vertical magnetic field with $B_z = 10^{-3}~\mu$G that provides a low strength magnetic background throughout the galactic disk and the low density background medium. The second component added a toroidal magnetic field inside the galactic disk with $B_\mathrm{tor} = 10^{-1} ~\mu\mathrm{G} \times \sqrt{\rho / \rho_\mathrm{max}}$, where $\rho$ is the gas density and the maximum gas density $\rho_\mathrm{max} = 2.06~m_\mathrm{p}~\mathrm{cm}^{-3}$ at the centre of the galaxy. The choice of the initial magnetic field strength in isolated disk galaxy simulations impacts its evolution, as pointed out by \citet{2025Kjellgren}. Our was chosen such that $v_\mathrm{a} = B(\rho) / \sqrt{4 \pi \rho} = \mathrm{const} \simeq 0.15~\mathrm{km}~\mathrm{s}^{-1}$ which is comparatively low compared to the sound speed and the rotation velocity.

Star formation was modelled using a standard Schmidt-type recipe, for which we use an efficiency per free fall time $\varepsilon_\mathrm{ff} = 100 \%$ above a hydrogen number density of $100~\mathrm{cm}^{-3}$. Stellar feedback yields for early stellar feedback and supernova feedback (in terms of energy, momentum, mass and metal returns) are based on \textsc{Starburst99} \citep{1999Leitherer} calculations. For our target mass, each star particle represents a single stellar population and hosts multiple SN type II. For each of those, we either injected $1.06 \times 10^{51}$ erg of thermal energy if the resulting blast wave is resolved or $4\times10^5~\mathrm{M}_\odot~\mathrm{km}~\mathrm{s}^{-1}$ of momentum if the SN explosion is unresolved. In the unresolved case, the ejecta momentum was injected into the surrounding medium in a manifestly isotropic fashion, which prevents biases introduced by simplistic mass-weighting \citep{2018Smith}. CR energy was injected for each SN event and we assumed an acceleration efficiency of $5\%$ \citep{2014Caprioli,2018Pais}.

The thermochemical state of the ISM was modelled using the non-equilibrium thermochemistry module of \textsc{Crisp}. It follows the abundances of molecular, atomic and ionized hydrogen, all ionization states of helium, as well as the first two ionization states of carbon, oxygen, and silicon. For these metals, we compute low-temperature cooling functions on-the-fly \citep[using rates from][]{2007Abrahamsson, 2014Grassi}, while high-temperature metal-line cooling is modelled from pre-calculated \textsc{Chianti} tables \citep{1997Dere}, assuming collisional ionization equilibrium. We further follow Ly$\alpha$ cooling of hydrogen \citep{1992Cen}, molecular hydrogen line cooling \citep{2021Moseley}, and bremsstrahlung cooling at high temperatures \citep{2018Ziegler}. These cooling channels are counteracted by photoelectric heating in the ISM \citep{1994Bakes}. The strength of photoelectric heating depends on the local far-UV photon flux, which we compute using a tree-based ray-tracing approach similar to that described by \citet{2021Wuensch}. We accounted for \citet{2013Rahmati}-attenuated photoionzation using the UV background of \citet{2019Puchwein}, collisional ionization \citep{1992Cen}, recombination \citep{1992Cen}, dust-assisted recombination \citep{2001Weingartner} and molecular hydrogen formation \citep{1979Hollenbach}, as well as charge exchange reactions \citep{2007Glover}. The union of these processes set the dynamical temperature and ionization state of the ISM -- a prerequisite to calculate the correct ion-neutral damping rates.

\section{CRs in the galactic environment}
\label{sec:overview}

\begin{figure*}
   \centering
   \resizebox{\hsize}{!} { \includegraphics[width=\textwidth]{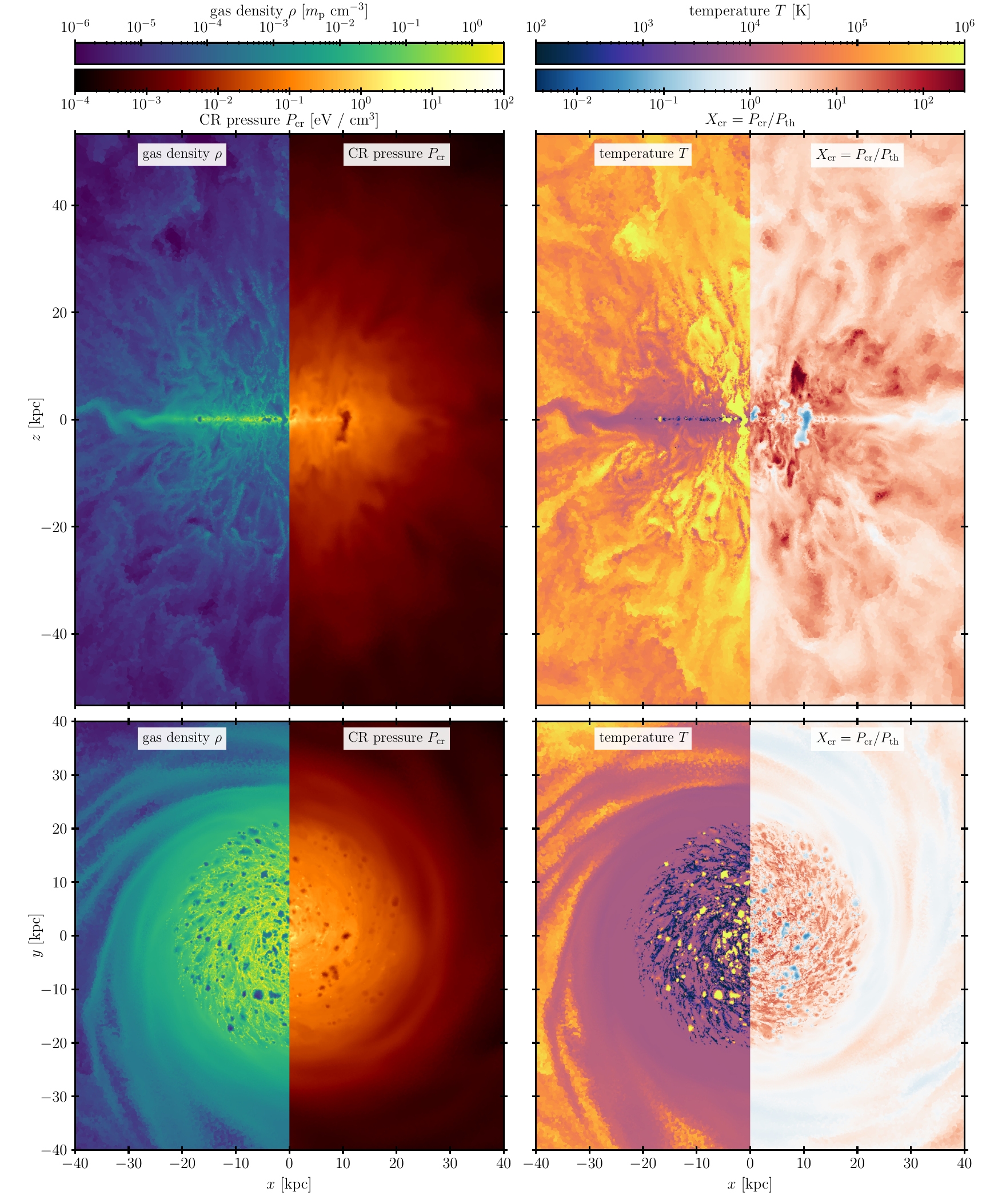} }
   \caption{Gallery showing edge-on slices through the galactic disk and wind (top row) and face-one slices through the galactic midplane (bottom row). Gas related quantities (density $\rho$ and temperature $T$) can be found on the left-hand side of each column, while CR related quantities are on the right-hand side: CR pressure $P_\mathrm{cr}$ and CR-to-thermal pressure ratio $X_\mathrm{cr} = P_\mathrm{cr} / P_\mathrm{th}$. The ISM and the CGM of the galaxy are highly structured and multiphase owning to stellar feedback. The CR population is relatively smooth and exhibits pressures exceeding the thermal pressure within the galactic wind.}
   \label{fig:gallery_1}
\end{figure*}

\begin{figure*}
   \centering
   \resizebox{\hsize}{!} { \includegraphics[width=\textwidth]{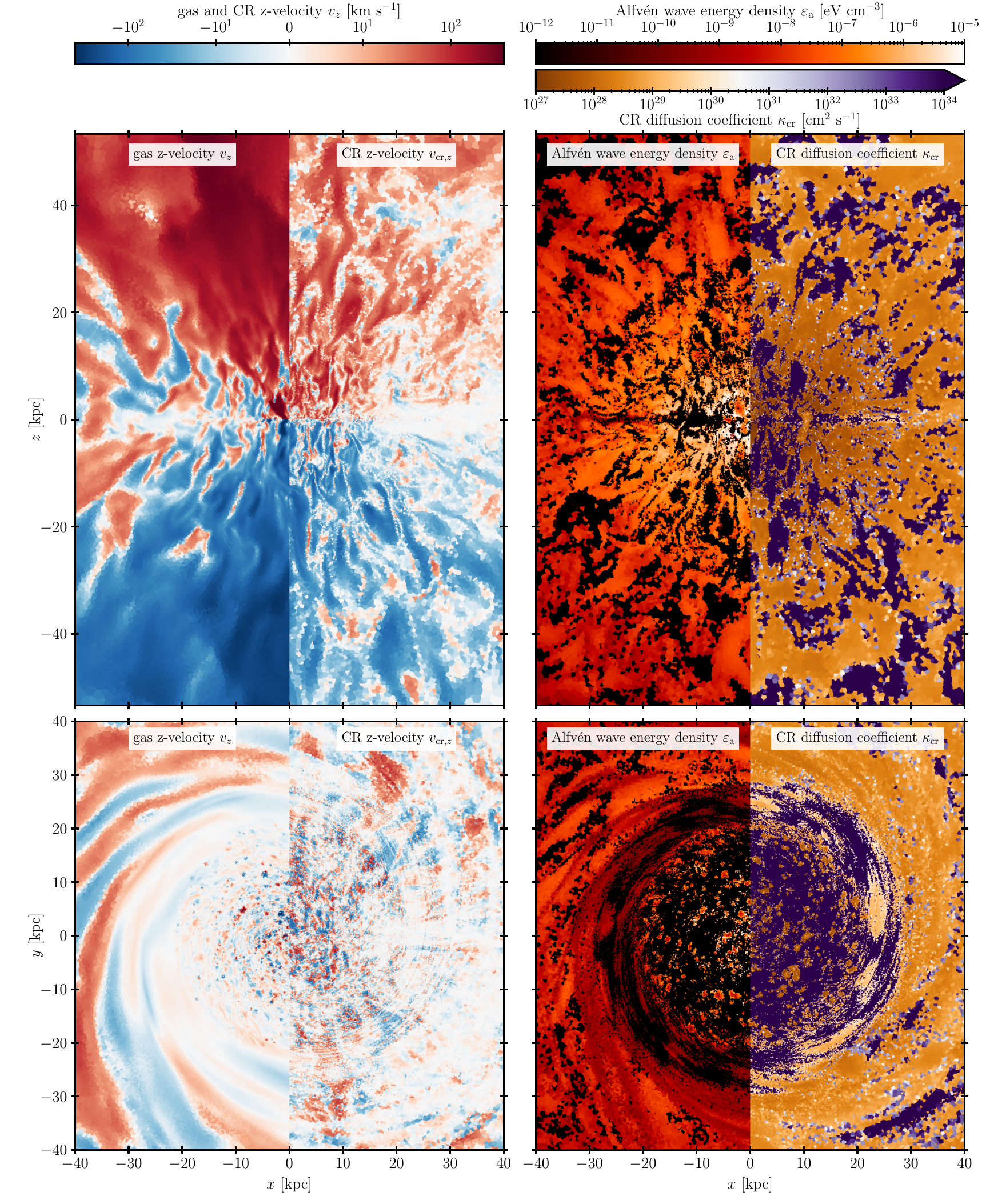} }
   \caption{Gallery similar to Fig.~\ref{fig:gallery_1} but showing the vertical gas velocity $\varv_{z}$ and the projected vertical CR transport speed along magnetic field lines $ \varv_{\mathrm{cr}, z}$ (left panels) and the Alfv\'en wave energy density $\varepsilon_\mathrm{a}$ and the CR diffusion coefficient $\kappa_\mathrm{cr}$ (right panels). Because CRs are advected with the gas, the total vertical CR transport speed is given by the sum of the gas and CR velocities. Both the Alfv\'en wave energy density $\varepsilon_\mathrm{a}$ and the CR diffusion coefficient $\kappa_\mathrm{cr}$ are highly structured, highlighting Alfv\'en-wave dark regions in the galactic outflow and an absence of Alfv\'en waves in the ISM with the exception of hot (super-) bubble regions. }
   \label{fig:gallery_2}
\end{figure*}

\begin{figure}
   \centering
   \resizebox{\hsize}{!} { \includegraphics[width=\textwidth]{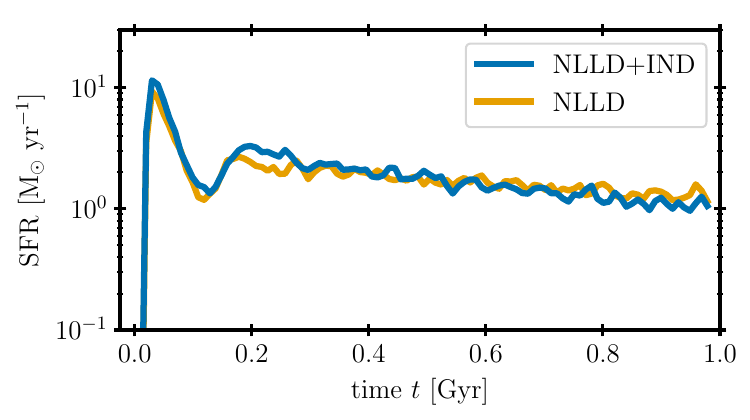} }
   \caption{Star formation rates for the two simulated CR-transport models. Both galaxies exhibit similar SFRs, with comparable starburst periods at the start of the simulations, followed by a brief phase of minor quenching.}
   \label{fig:star_formation_rate}
\end{figure}

\begin{figure}
   \centering
\resizebox{\hsize}{!} { \includegraphics[width=\textwidth]{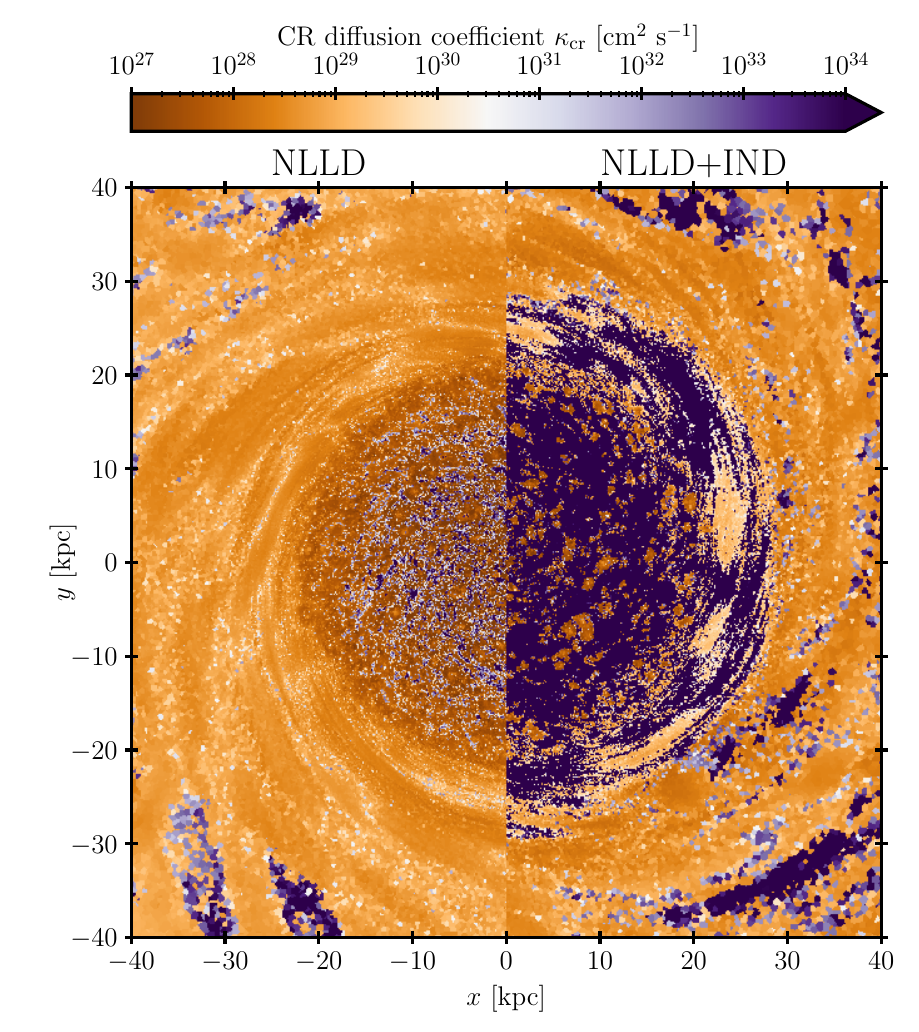} }
   \caption{Gallery similar to Fig.~\ref{fig:gallery_2} but showing the CR diffusion coefficient $\kappa_\mathrm{cr}$ for both the NLLD (left half) and the NLLD+IND (right half) simulation as slices through the galactic plane. The inclusion of ion–neutral damping in the NLLD+IND simulation reduces the abundance of CR-scattering Alfvén waves in the ISM, resulting in a correspondingly higher diffusion coefficient. }
   \label{fig:gallery_kappa_comparision}
\end{figure}

Before analysing the details of CR transport inside the simulations and its influence on the galactic-scale properties of the CR population, we first use this section to provide an overview of the simulated Milky Way-like NLLD+IND galaxy. 

In Fig.~\ref{fig:gallery_1}, we display various hydrodynamical quantities using slices through the $xz$-plane (showing the galactic disk edge-on as well as the galactic wind) and $xy$-plane (showing a face-on slice through the galactic disk). Both the galactic wind and the ISM of the galaxy are highly structured as a consequence of stellar feedback and a CR-driven outflow. Inside the galactic disk, energy and momentum injected by clustered SNe drive underdense ($\rho < 10^{-3} \, m_\mathrm{p}\, \mathrm{cm}^3$) and hot ($T > 10^{5}$ K) (super-) bubbles into their surrounding ISM. The warm ($T \sim 10^{4}$ K) ISM builds up a diffuse gaseous layer that enshrouds the filamentary cold ($T < 10^{4}$ K) ISM. Various cooling processes cause the emergence of the cold phase, giving eventually rise to molecular clouds that collapse, fragment and ultimately form a new generation of stars. Stellar feedback from SNe initiates the galactic wind-driving process by launching gas above the galactic disk. There, the gas can be found in an intermediate state between a fountain flow, where dense gas is expelled from the disk but later falls back, and a smooth galactic outflow that is driven away from the galaxy. The galactic wind has an intriguing morphology with dense filamentary or scale-like features filling the inner CGM which are surrounded by diffuse and lower density gas. The presence of these dense features is unique to CR-driven galactic winds \citep{2016Girichidis, 2025Thomas} where the additional CR-pressure inside the inner CGM supports these gaseous structures. Visually correlating gas densities and temperatures shows that the dense features correspond to warm gas, while the diffuse low density gas is hot.

Figure~\ref{fig:gallery_1} also shows the CR pressure $P_\mathrm{cr}$ and the CR pressure-to-thermal pressure ratio $X_\mathrm{cr} = P_\mathrm{cr} / P_\mathrm{th}$, which indicates whether CRs are overpressurized compared to the thermal gas. Overall, the CR-pressure distribution is rather smooth throughout the galactic disk and the inner CGM. Inside the ISM, CR-pressure is enhanced around sites of recently injected stellar feedback, where CRs are freshly injected by young SN remnants. The $P_\mathrm{cr}$ distribution has holes in expanding (super-)bubble structures where CRs are adiabatically expanded. Outside the galactic disk and within the galactic wind, the CR pressure follows the structure of the gas density but is considerably smoother, lacking strong variations between regions of warm-dense and hot-dilute gas. There is a global CR-pressure gradient pointing toward the galactic disk, which has a high CR-pressure ($P_\mathrm{cr}\sim 1\,\mathrm{eV}\,\mathrm{cm}^{-3}$), into the CGM, which has lower CR pressures ranging over multiple orders of magnitude. This is expected as the source of CRs in our simulation is stellar feedback, which itself is situated in the galactic disk, and CRs lose energy via adiabatic, cooling, and transport processes on their way from the ISM into the CGM. The inner CGM is overpressurized in CRs relative to the thermal pressure, with $X_\mathrm{cr} \sim 1$–$10$, reaching values of up to $\sim$100 in a few regions. Inside the ISM, the CR pressure also exceeds the thermal pressure with the notable exceptions of expanding (super-)bubbles where the thermal pressure dominates. In the outskirts of the gaseous galactic disk, no in-situ star formation takes place and CRs reach these regions only via transport. In such regions, the thermal pressure can surpass the CR pressure because only a negligible fraction of CRs propagates into them.

In Fig.~\ref{fig:gallery_2} we provide an additional gallery of figures which are related to the transport of CRs. The panels on the left-hand side show the vertical gas velocity $\varv_z$ and vertical CR transport speed $\varv_{\mathrm{cr}, z} = (\bs{e}_z \bcdot \bs{b}) \varv_\mathrm{\mathrm{cr}}$, where the projection of the direction of the magnetic field $\bs{b} = \bs{B} / B$ onto the $z$-axis describes how fast CR energy is transported along magnetic fields in the vertical direction. The gas velocity is turbulent close to the galactic disk, where fountain flows are actively redistributing gas that has been launched out of the ISM via stellar feedback. Above this turbulent layer, a clear biconical outflow can be observed, which locally reaches speeds of up to $300~\mathrm{km}~\mathrm{s}^{-1}$. 

The vertical CR transport speed also reaches nearly $300~\mathrm{km}~\mathrm{s}^{-1}$ but is generally slower than the gas velocity. The vertical transport speed of CRs is determined by the sum of $\varv_z$ and $\varv_{\mathrm{cr}, z}$, where $\varv_{\mathrm{cr}, z}$ tells us the CR transport speed along magnetic field lines while $\varv_z$ shows the advection speed of these field lines. Consequently, $\varv_{\mathrm{cr}, z}$ can be seen as the differential vertical transport speed of CRs. Inside the galactic wind, it has more structure than its gas counterpart and has frequent sign changes. Because of its differential nature, this does not directly imply that there are backflows of CRs towards the galactic disk. 

Visual artifacts of the vertical CR transport velocity can be seen inside the galactic disk. Because the magnetic field inside the disk is highly toroidal, slight fluctuations in the vertical magnetic field direction cause sign changes of $\varv_{\mathrm{cr},z}$ because it is essentially a vertical projection of the magnetic field direction. Furthermore, the CR transport velocity $\varv_\mathrm{cr}$ is also a signed quantity, telling us whether CRs are transported in the direction pointing along the magnetic field line or along the directly opposite direction, which adds another possibility to switch the sign of $\varv_{\mathrm{cr},z}$. 

The right-hand panel of Fig.~\ref{fig:gallery_2} shows the Alfv\'en wave energy density $\varepsilon_\mathrm{a}$ and the CR diffusion coefficient $\kappa_\mathrm{cr}$. Both quantities are related to each other via $\kappa_\mathrm{cr} \propto \varepsilon_\mathrm{a}^{-1}$, explaining the similar structures in both quantities. The most striking features are the Alfv\'en-wave dark regions inside the outflow, where Alfv\'en wave energy is vanishingly low and the CR diffusion coefficient is high, and the similar absence of Alfv\'en waves and the accompanying high CR diffusion coefficient inside the galactic disk. The latter appears to be almost disk like and traces the ISM of the galaxy if compared to the presence of cold and warm gas in Fig.~\ref{fig:gallery_1}. This disk-like structure is only interrupted by bubble-like features, which correspond to stellar feedback-affected hot (super-)bubble regions, as can be verified by locating them in the temperature field in Fig.~\ref{fig:gallery_1}. We will show below that these regions of large intrinsic $\kappa_\mathrm{cr}$ are caused by ion-neutral damping.

Back to Alfv\'en-wave dark regions, which were shown to arise where CRs propagate more slowly than the Alfv\'en waves, causing the waves' magnetic energy to be efficiently converted into CR kinetic energy \citep{2023Thomas}. This is the exact opposite (i.e., the time-reversed process) of the streaming instability, where CRs move faster than the Alfv\'en waves and their kinetic energy is used to amplify the waves' magnetic energy. For these regions to appear, the CR transport speed along magnetic field lines needs to be slower than the local Alfv\'en speed. This might happen in regions where the CRs propagate slowly in general and $\varv_\mathrm{cr}$ is close to zero. At such positions, $\varv_\mathrm{cr}$ and, consequently, $\varv_{\mathrm{cr}, z}$ may switch signs. Comparing the positions where $\varv_{\mathrm{cr}, z}$ switches signs in the galactic wind in the left-hand panel of Fig.~\ref{fig:gallery_2} to the positions of Alfv\'en-wave dark regions in the right-hand panel, shows that they are visually correlated. 

In the following, we compare the results of the NLLD and NLLD+IND simulation runs. Both galaxies should show a similar evolution over the course of the simulation to facilitate a fair comparison. In Fig.~\ref{fig:star_formation_rate}, we show the star formation rate (SFR) of both galaxies and find that they are similar to the extent that even the level of temporal fluctuations in both simulations is the same. Also, the initial starburst phase and the following short period of minor star formation quenching are quite comparable between the two runs. This result is expected because \citet{2025Thomas} demonstrated that including CR feedback has only a minor impact on the SFR in our idealized setup; thus, altering CR propagation properties in the ISM likewise has minimal influence on the SFR.

In Fig.~\ref{fig:gallery_kappa_comparision}, we compare the CR diffusion coefficient in a slice through the galactic disk for the NLLD and NLLD+IND simulation runs, similar to Figs.~\ref{fig:gallery_1} and \ref{fig:gallery_2}. While the NLLD+IND simulation has high CR diffusion coefficients throughout the galactic disk with exceptions of the localized feedback-affected bubble regions, the NLLD simulation has mostly low CR diffusion coefficients inside the ISM. In this simulation, only small patches of the ISM have high CR diffusion coefficients. As no ion-neutral damping is operating in this simulation, only NLLD and Alfv\'en-wave dark regions can regulate CR transport. Since NLLD does not fully damp Alfv\'en waves but instead caps their growth at levels corresponding to low CR diffusion coefficients, the resulting localized patches manifest as Alfv\'en-wave dark regions in the ISM. The existence of such regions is also possible in the NLLD+IND simulation, but they are overshadowed by the large-scale impact of ion-neutral damping on Alfv\'en-wave dynamics in the ISM.

\begin{figure*}
   \centering
   \resizebox{\hsize}{!} { \includegraphics[width=\textwidth]{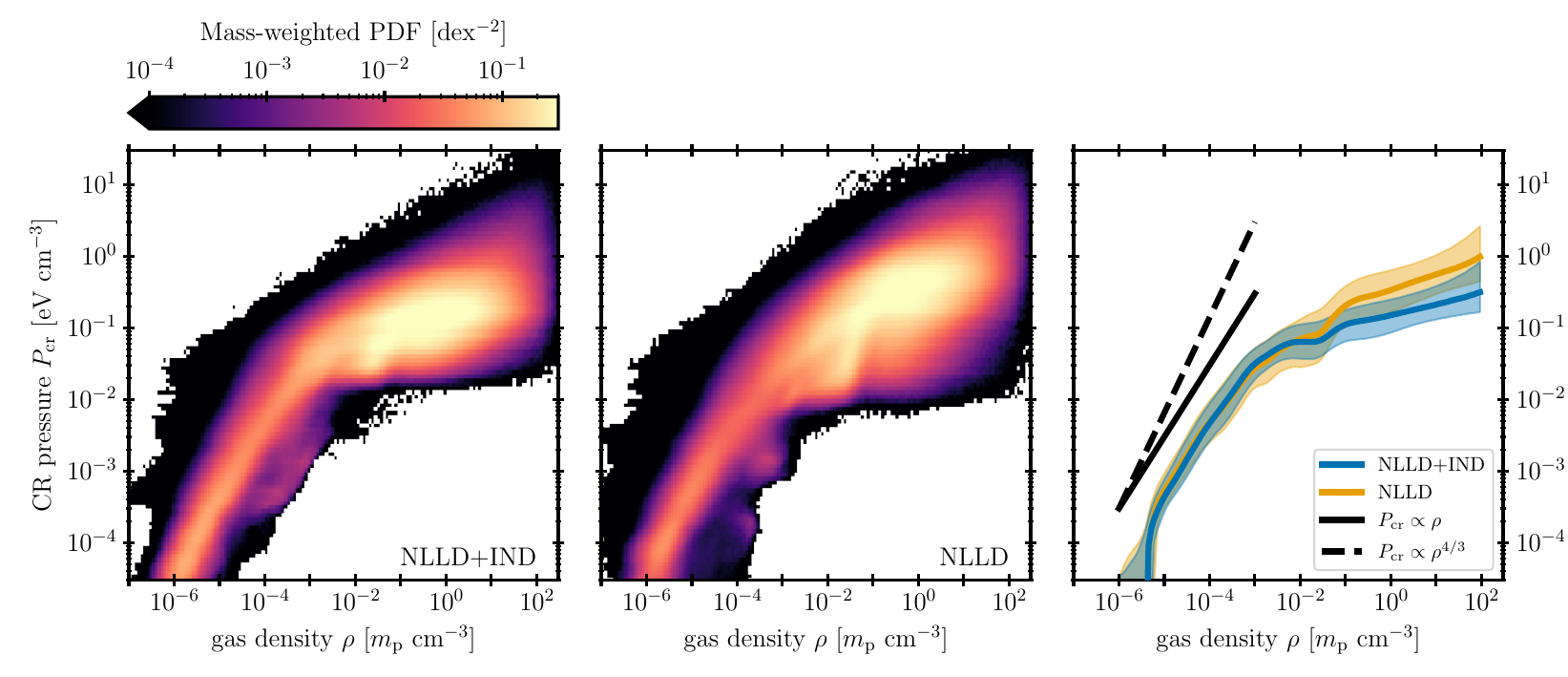} }
   \caption{ CR-phase diagram relating CR pressure $P_\mathrm{cr}$ and gas density $\rho$ for both the NLLD+IND (left panel) and the NLLD simulation (middle panel) displayed as a mass-weighted PDF taking into account data of the last 100~Myr of evolution from the star-forming disks and galactic outflows. In the right panel, we show the median as a solid line and 20$^\mathrm{th}$ and 80$^\mathrm{th}$ percentiles as shaded regions. Both simulations show different relationships between $\rho$ and $P_\mathrm{cr}$ in dilute and dense media with a prominent transition region at around $\rho \sim 10^{-2}~m_\mathrm{p}~\mathrm{cm}^{-3}$. At high densities, the NLLD run shows systematically increased CR pressures because of the slow CR diffusion in this medium which leads to long CR escape times. }
   \label{fig:rho_Pcr_correlation}
\end{figure*}

\begin{figure*}
   \centering
   \resizebox{\hsize}{!} { \includegraphics[width=\textwidth]{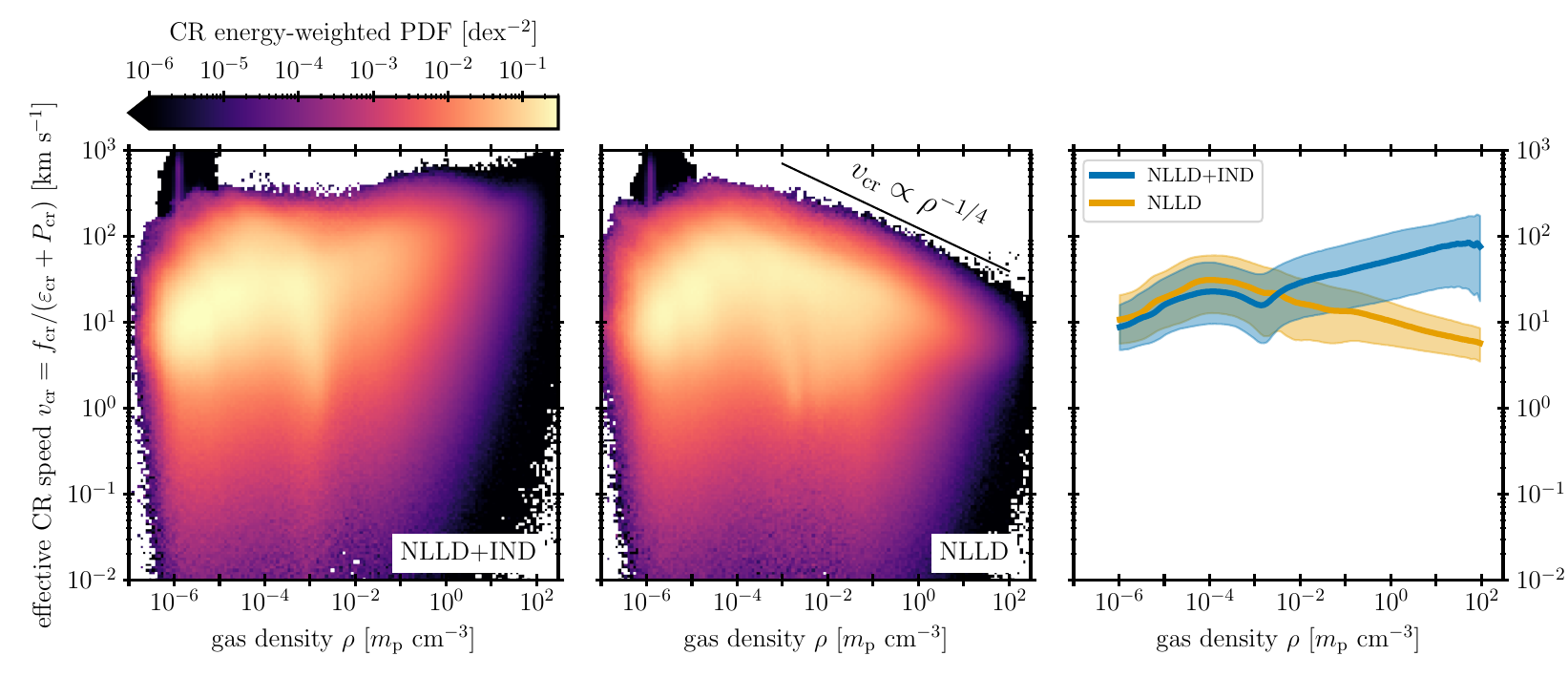} }
   \caption{Distribution of the effective CR transport speed as a function of the gas density for both the NLLD+IND (left panel) and the NLLD simulation (middle panel) displayed as a CR energy-weighted PDF considering data of the last 100~Myr of evolution. In the right panel, we show the median as a solid line and 20$^\mathrm{th}$ and 80$^\mathrm{th}$ percentiles as shaded regions. In the middle panel, we highlight the enveloping relation $\varv_\mathrm{cr} \sim \rho^{-1/4}$, where the power-law index is fitted by hand. Both simulations show widely different CR transport speeds at higher densities.}
   \label{fig:rho_vcr_correlation}
\end{figure*}

\section{Effective Transport}
\label{sec:effective CR transport}

In this section, we characterize CR transport through its effective transport parameters and distinguish primarily between the cold-to-warm ISM and the warm-to-hot galactic outflow. Because the main difference between the two simulations is ion-neutral damping, which operates only in the presence of neutral particles, we expect its largest impact to appear in the ISM. To motivate that CR transport has a tangible impact in our simulations, we show a density-pressure diagram for the CR pressure $P_\mathrm{cr}$ in Fig.~\ref{fig:rho_Pcr_correlation}. For the presented probability distribution function (PDF), we exclude cells that are within the extended \ion{H}{I} disk in our simulation. The \ion{H}{I} disk is not magnetically well-connected to the outflow and the active ISM and hence is ineffectively populated by CRs. Its inclusion introduces features (which are still somewhat visible at densities around $\rho \sim 3\times10^{-2}~m_\mathrm{p}\,\mathrm{cm}^{-3}$) that would distract from the otherwise clear picture drawn in Fig.~\ref{fig:rho_Pcr_correlation}. We include computational cells that are with a (cylindrical) radius of $R < 20~\mathrm{kpc}$ irrespective of height and include all CGM cells above of height of $|z| > 2~\mathrm{kpc}$ and use data from the last 100 Myr of evolution. Thus, these criteria include the ISM and the galactic outflow. Both simulations show a unimodal distribution in the $\rho$ -- $P_\mathrm{cr}$ plane where the correlation between the two quantities is rather tight at low densities and broadens at higher densities. At low densities, the correlation appears to be power law-like and roughly follows $P_\mathrm{cr} \propto \rho$. The corresponding power-law index of one is in between $4/3$, which expected for a purely adiabatic CR fluid, and $2/3$ for a purely streaming CR fluid in a flux-rope geometry \citep{1991Breitschwerdt}. This indicates that both, adiabatic and streaming processes affect CR transport at low density. This is in line with the findings of \citet{2025Sampson}, where the authors investigated CR-transport in driven isothermal turbulence and showed that their $\rho$ -- $P_\mathrm{cr}$ distribution is bound by these two power laws in the case of trans- to super-Alfv\'enic turbulence (which is representative of the low density CGM). Both, the NLLD and NLLD+IND simulations have almost identical mass-weighted medians and 20$^\mathrm{th}$ to 80$^\mathrm{th}$ percentile curves, which indicates that their transport is independent of ion-neutral damping once CRs left the high-density region of the ISM. These curves differ at densities above $\rho > 10^{-3}~m_\mathrm{p}\,\mathrm{cm}^{-3}$ and the NLLD simulation shows a systematic increase of $P_\mathrm{cr}$ up to a factor of three at the highest densities reached in our simulation. Also, the distribution of the NLLD simulations is broader compared to the NLLD+IND simulation at these densities. Interestingly, both simulations have virtually identical SFRs, which indicates that the increased CR pressures at high densities (close to the star formation threshold) have only a minor impact on the simulated dynamics that ultimately lead to star formation. Investigating whether and under which conditions CRs impact the star formation process at the molecular cloud-scale level is beyond the scope of this work. 

Owing to the similar SFRs, CRs are also injected into the ISM at comparable rates, the difference in CR transport is the only viable explanation for the enhanced CR pressures in the ISM seen in the NLLD simulation. To check this hypothesis, we show in Fig.~\ref{fig:rho_vcr_correlation} the PDF of gas densities $\rho$ and effective CR transport speed $\varv_\mathrm{cr}$. We use data points from the last 100 Myr of evolution and employ a CR energy weighting to present the statistics from the viewpoint of the CR population. The bulk of the CR energy is transported with speeds ranging between $10~\mathrm{km}\,\mathrm{s}^{-1}$ and $100~\mathrm{km}\,\mathrm{s}^{-1}$. At the transition density $\sim 10^{-3}~m_\mathrm{p}\,\mathrm{cm}^{-3}$, a small feature that extends the bulk of the distribution towards lower $\varv_\mathrm{cr}$ can be observed. Below this transition density, both simulations have similar transport speeds at similar densities. As both simulation share the same physical model at low densities, this is expected. At higher densities, CR energy in the NLLD+IND simulation is transported faster, while in the NLLD simulation it is transported more slowly compared to the common low-density regime. For the NLLD simulation, we show a $\varv_\mathrm{cr} \propto \rho^{-1/4}$ power law that fits the envelope of the high density distribution. A simple explanation for this power law could be inferred by assuming streaming CRs ($\varv_\mathrm{cr} \sim \varv_\mathrm{a,i}$) and spherically compressed magnetic fields ($B \propto \rho^{2/3}$) but this exercise results in $\varv_\mathrm{cr} \propto \rho^{+1/6}$ which yields the opposite trend and leads to faster CR transport at higher density.
The failure of this simple explanation hints towards additional operating physics. Having established that CRs are faster inside the ISM in the NLLD+IND simulation compared to their NLLD counterparts, the increased CR pressure in the NLLD simulation can be explained by a longer residency time of CRs in the NLLD simulation.

\begin{figure*}
   \centering
   \resizebox{\hsize}{!} { \includegraphics[width=\textwidth]{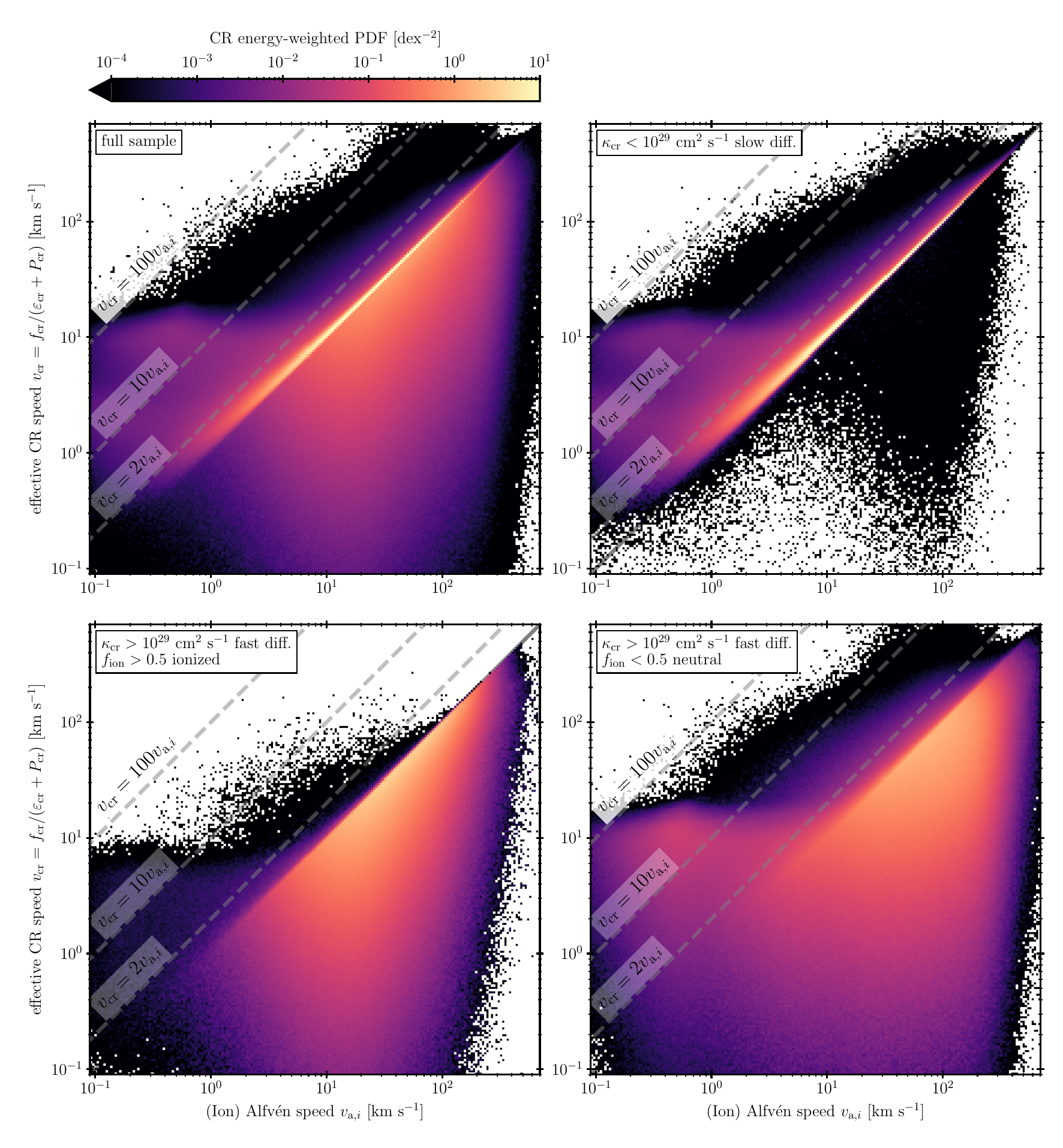} }
   \caption{CR energy-weighted distribution of the effective CR transport speed as a function of the (ion) Alfv\'en speed $\varv_{\mathrm{a}, i}$ for the NLLD+IND simulation. We separately display all analysed data points (top-left), data points that are in regions where intrinsic CR diffusion is slow and the coupling between CRs and Alfv\'en waves is strong, $\kappa_\mathrm{cr} < 10^{29}~\mathrm{cm}^2~\mathrm{s}^{-1}$, (top-right) and regions where CRs are uncoupled ($\kappa_\mathrm{cr} > 10^{29}~\mathrm{cm}^2~\mathrm{s}^{-1}$) and their surrounding environment is either ionized (bottom-left) or neutral (bottom-right). While most CRs are coupled and stream at the Alfv\'en speed, CRs that are uncoupled are preferentially slower than the Alfv\'en speed or, if they stream with super-alfven\'ic speeds, they are in neutral environments.}
   \label{fig:cosmic_ray_speed}
\end{figure*}

To put the effective transport speeds into context, we compare $\varv_\mathrm{cr}$ with the local (ion) Alfv\'en speed $\varv_{\mathrm{a}, i}$ in Fig.~\ref{fig:cosmic_ray_speed} where we show a CR-energy weighted PDF calculated from the last 100 Myr of evolution from the NLLD+IND simulation. Through the self-regulation of CR transport via the streaming stability, most of the CRs can be found to be streaming with the (ion) Alfv\'en speed and have $\varv_\mathrm{cr} \sim \varv_{\mathrm{a}, i}$ which is notably visible in Fig.~\ref{fig:cosmic_ray_speed} as the accumulation of CR energy on the diagonal. In order to discuss deviations from this relation, we show the PDFs for CRs that are well coupled, with a low intrinsic diffusion coefficient ($\kappa_\mathrm{cr} < 10^{29}~\mathrm{cm}^2~\mathrm{s}^{-1}$), and for CRs that are uncoupled, with a high intrinsic diffusion coefficient ($\kappa_\mathrm{cr} > 10^{29}~\mathrm{cm}^2~\mathrm{s}^{-1}$). A low intrinsic diffusion coefficient implies that CRs are efficiently scattered, such that their transport speeds will be close to the Alfv\'en speeds. These CRs are rarely slower than the Alfv\'en waves because they need to be faster than $\varv_{\mathrm{a}, i}$ to excite the streaming instability, allow for a growth of Alfv\'en waves, which will eventually cause the low intrinsic diffusion coefficient in the first place. The majority of the streaming CRs fall into this category. Some CRs that are contained within regions where $\varv_{\mathrm{a}, i} \lesssim 3~\mathrm{km}~\mathrm{s}^{-1}$ can be found to be streaming at highly super-Alfv\'enic speeds. Because the energy transfer rate scales linearly with $\varv_{\mathrm{a}, i}$, see Eq.~\eqref{eq:ealf}, which is low in these regions, these CRs have prolonged timescales on which scattering Alfv\'en waves could possibly be excited. This leaves the CRs unscattered and allows them to stream super-Alfv\'enically. To analyse CRs with high intrinsic diffusion coefficients, we show separate PDFs for CRs in ionized ($f_\mathrm{ion} > 0.5$) and neutral environments ($f_\mathrm{ion} < 0.5$). CRs in ionized regions with high intrinsic diffusion coefficients can be spatially found in Alfv\'en wave dark regions, where the CRs extract energy from the Alfv\'en wave pool if they are slower than the (ion) Alfv\'en speed. Their corresponding PDF in Fig.~\ref{fig:cosmic_ray_speed} is consequently dominantly populated below the $\varv_\mathrm{cr} \sim \varv_{\mathrm{a}, i}$ diagonal. Similar populations can also be found in neutral regions, with some additional CRs being located above the diagonal. Their presence in this parameter space is explained by ion-neutral damping. Although CRs should excite Alfv\'en waves through the gyroresonant streaming instability, their growth is suppressed by ion–neutral damping, so that no scattering Alfv\'en waves build up, resulting in a high intrinsic CR diffusion coefficient.

\section{Effective Diffusion}

\begin{figure*}
   \centering
   \resizebox{\hsize}{!} { \includegraphics[width=\textwidth]{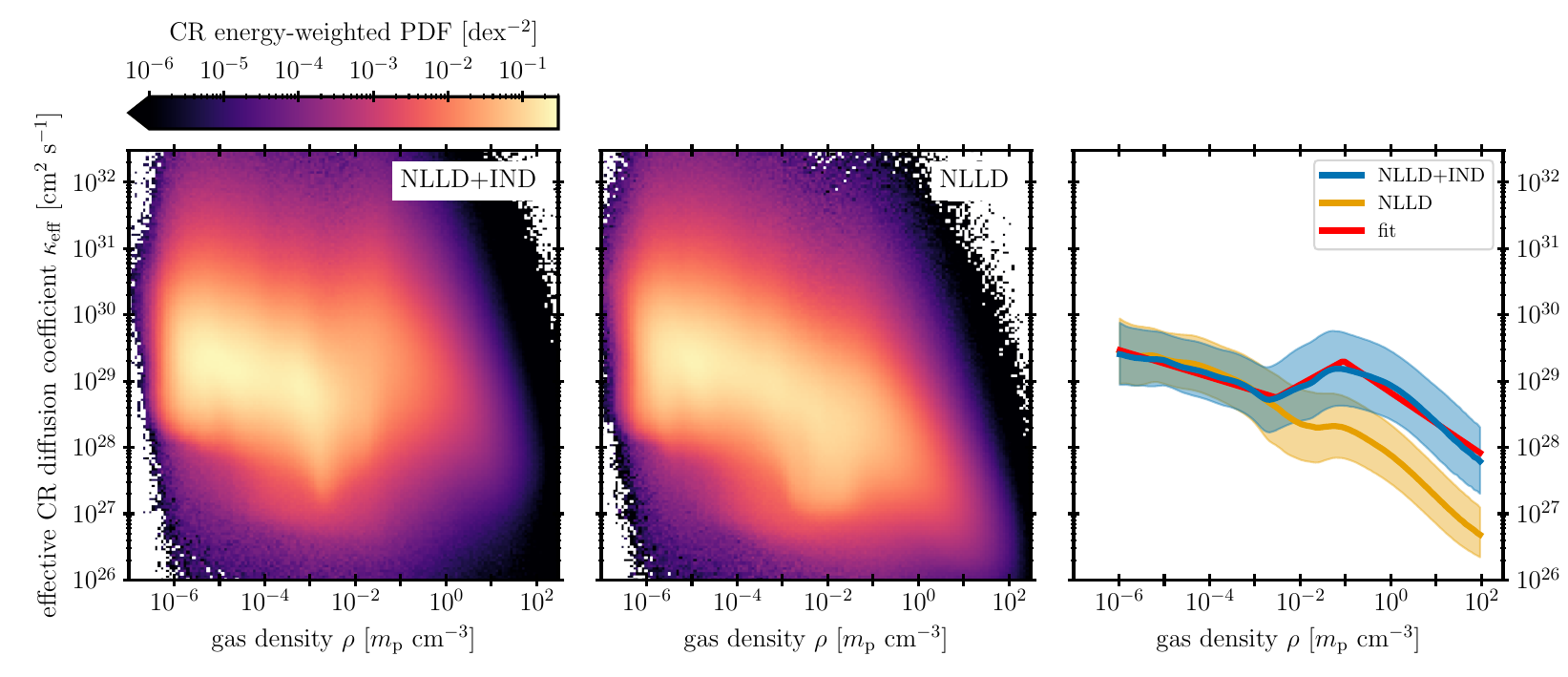} }
   \caption{ Distribution of the effective CR diffusion coefficient $\kappa_\mathrm{eff}$ as a function of the gas density for both the NLLD+IND (left panel) and the NLLD simulation (middle panel) displayed as a CR energy-weighted PDF considering data of the last 100~Myr of evolution. In the right panel, we show the median as a solid line and 20$^\mathrm{th}$ and 80$^\mathrm{th}$ percentiles as shaded regions and a fit to the NLLD+IND median curve. Both simulations have similar distributions at lower densities but differ inside denser environments where neutral atoms coexists with ions and ion-neutral damping starts to affect CR transport.}
   \label{fig:rho_kappa_eff_correlation}
\end{figure*}

\begin{figure*}
   \centering
   \resizebox{\hsize}{!} { \includegraphics[width=\textwidth]{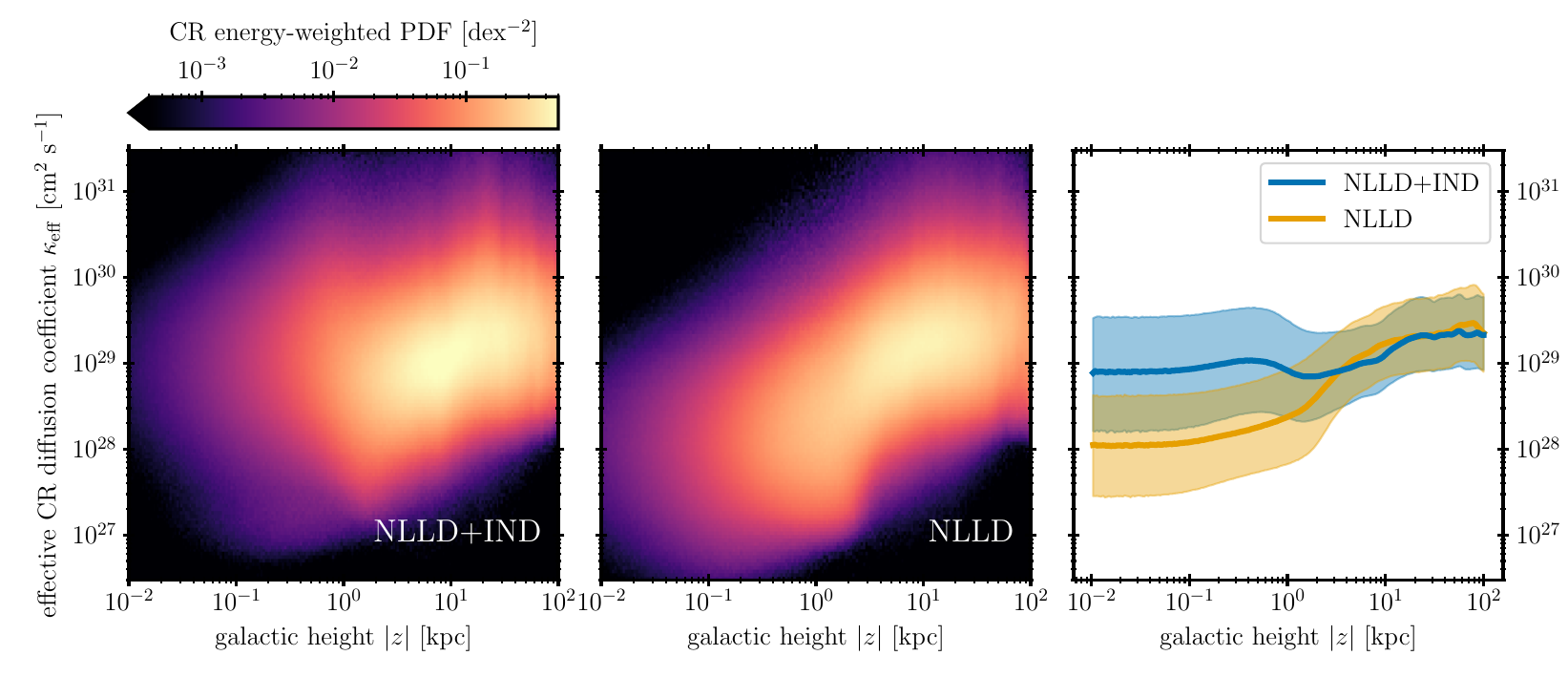} }
   \caption{Distribution of the effective CR diffusion coefficient $\kappa_\mathrm{eff}$ (defined in Eq.~\ref{eq:kappa_eff}) as a function of the galactic height for both the NLLD+IND (left panel) and the NLLD simulation (middle panel) displayed as a CR-energy-weighted PDF considering data of the last 100 Myr of evolution. In the right panel, we show the median as a solid line and 20$^\mathrm{th}$ and 80$^\mathrm{th}$ percentiles as shaded regions. Both simulations have similar distributions of $\kappa_\mathrm{eff}$ at larger galactic heights but differ at low galactic heights and hence inside the ISM.}
   \label{fig:vertical_distribution}
\end{figure*}

Alternatively, we can reanalyse the previous statements in terms of the effective diffusion coefficient. The effective transport speed and the effective diffusion coefficient are related via:
\begin{equation}
    \kappa_\mathrm{eff} = - L_\mathrm{cr} \,\varv_\mathrm{cr} \gamma_\mathrm{cr},
\end{equation}
where $L_\mathrm{cr} = \varepsilon_\mathrm{cr} / \bs{b} \bcdot \bnabla \varepsilon_\mathrm{cr}$ is the projected scale-length of the CR energy density.\footnote{Note that in contrast to the usual definition of a length scale, $L_\mathrm{cr}$ is a signed quantity.} In Fig.~\ref{fig:rho_kappa_eff_correlation}, we show the PDF of the effective diffusion coefficient $\kappa_\mathrm{cr}$ versus gas density $\rho$. We, again, display the results for both simulations using data from the last 100 Myr of evolution and show the median and the 20$^\mathrm{th}$ and 80$^\mathrm{th}$ percentiles of the CR energy-weighted statistics. Overall, the distributions appear to be broader as those for the effective transport speed $\varv_\mathrm{cr}$ in Fig.~\ref{fig:rho_vcr_correlation} and span multiple orders of magnitude. This impression is mostly caused by the tails of the distribution at a given gas density because most CRs can still be found inside a band of effective CR diffusion coefficients that only ranges over a single order of magnitude. Similarly to our previous discussion, both distributions are similar at low densities, where both CR transport models do not differ. Consequently, the median curves also trace each other at densities below $\rho \sim 10^{-3}~m_\mathrm{p}\,\mathrm{cm}^{-3}$. Above this density threshold, the gaseous media are affected by the lack of an ionization source and ion neutral damping starts to operate. This causes an increase in the effective CR diffusion coefficient at high densities for the NLLD+IND simulation where we also observed an increase in the effective CR transport in Fig.~\ref{fig:rho_vcr_correlation}. In this region, $\kappa_\mathrm{cr}$ is higher by a factor $\sim$10 in the NLLD+IND simulation in comparison to the NLLD simulation. The median curves roughly follow broken power laws at low and high densities in  both simulations. We provide a fit for the median effective CR diffusion coefficient of the NLLD+IND simulation in the form of:
\begin{align}
    \kappa_\mathrm{eff} = \begin{cases}
        \kappa_0 \left(\frac{\rho}{\rho_{0}}\right)^{\alpha_0}, & \phantom{\rho_0 < }\,\, \rho < \rho_{0}  \\
        \kappa_0 \left(\frac{\rho}{\rho_{0}}\right)^{[\log(\kappa_1) - \log(\kappa_0)] / [ \log(\rho_{1}) -  \log(\rho_{0})]}, & \rho_{0} < \rho < \rho_{1} \\
        \kappa_1 \left(\frac{\rho}{\rho_{ 1}}\right)^{\alpha_1}, & \rho_{1} < \rho
    \end{cases}
    \label{eq:kappa_eff_fit}
\end{align}
which is composed of power laws at high and low densities and a third connecting power at intermediate densities, which is introduced to ensure a continuous transition between the two extreme regimes. The fitted parameters are:
\begin{align}
    \rho_{0} &= 3\times10^{-3}~m_\mathrm{p}\,\mathrm{cm}^{-3} \nonumber \\
    \rho_{1} &= 9\times10^{-2}~m_\mathrm{p}\,\mathrm{cm}^{-3} \nonumber \\
    \kappa_0 &= 5.94 \times 10^{28} ~\mathrm{cm}^{2} ~\mathrm{s}^{-1} \nonumber \\
    \kappa_1 &= 1.99 \times 10^{29} ~\mathrm{cm}^{2} ~\mathrm{s}^{-1} \nonumber \\
    \alpha_0 &= -0.183 \nonumber \\
    \alpha_1 &= -0.457,
    \label{eq:kappa_eff_fit_param}
\end{align}
where the characteristic densities $\rho_0$ and  $\rho_1$ are chosen manually, while the other parameters are fitted to minimize deviations from the media curve.

A parametrization of the effective CR diffusion coefficient in this form can be used as an input for numerical one-moment CR diffusion solvers and captures the multifacet nature of the present two-moment Alfv\'en wave regulated CR transport in the multiphase galactic environments in a trackable manner. We decided to provide the fit as a function of density for the following reason: a parametrization independent of the CR energy or pressure is beneficial because doing so, as in $\kappa_\mathrm{eff} = \kappa_\mathrm{eff}(\varepsilon_\mathrm{cr})$, would turn the CR diffusion problem into a non-linear diffusion problem, which is challenging to solve numerically. Using other variables such as temperature $T$ or magnetic field strength $B$ resulted in $\kappa_\mathrm{eff}$-distributions which did not show good correlations between the variable and $\kappa_\mathrm{eff}$. Although multiple parameterizations were feasible, we selected the one exhibiting the strongest correlation.

In Fig.~\ref{fig:vertical_distribution}, we show the vertical distribution of the effective diffusion coefficient $\kappa_\mathrm{eff}$ using an CR energy-weighted PDF and include data points from the last 100 Myr of evolution. Because of the CR energy-weighting and the rather smooth spatial distribution of CRs inside the galactic disk, the shown distribution can also be interpreted to be close to a volume-weighted PDF. Correspondingly, this plot is implicitly biased in favour of CR transport in the warm and hot phases at a given height. Inside the galactic wind and thus at larger galactic heights ($|z| \gtrsim 3~\mathrm{kpc}$), both distribution are similar and their median curves lay almost on top of each other. At such heights, no neutrals are present in our simulations, implying absent ion-neutral damping and that the CR transport physics is uniquely dictated by the NLLD damping process. The resolution of our simulations is insufficient to capture all cold clouds entrained in the galactic wind, and we may therefore miss ion-neutral damping within the clouds. But the volume filling fraction of clouds (as formed by condensation through thermal instability) is found to be at the percent level in high-resolution idealized simulations of the CR-affected CGM \citep{2025Weber}. Consequently, their contribution can be neglected for most of the volume and the majority of CRs inside the galactic wind. Inside the ISM and thus at lower galactic heights, the distributions of $\kappa_\mathrm{eff}$ differ due to the presence of the cold phase inside the ISM. Because of the overall faster CR transport in the ion-neutral damping simulation, the effective CR diffusion coefficient in this region is larger than it's NLLD counterpart. As for Fig.~\ref{fig:rho_kappa_eff_correlation}, the median $\kappa_\mathrm{eff}$ values differ by roughly an order of magnitude and their curves level off at heights below 100 pc -- the NLLD simulation at $\sim$$10^{28} ~ \mathrm{cm}^2 \, \mathrm{s}^{-1}$ and the NLLD+IND simulation $\sim$$10^{29} ~ \mathrm{cm}^2 \, \mathrm{s}^{-1}$. As expected, these do not correspond to the values reached at the highest densities but rather those at $\rho \sim 3\times 10^{-1} ~m_\mathrm{p}\,\mathrm{cm}^{-3}$. This is not necessarily a statement that this is a typical density for CR transport, but rather exemplifies that CR transport inside the ISM takes places in a multiphase environment with highly varying ambient densities.

As already hypothesized by \citet{1969Kulsrud}, at high-densities, ion-neutral damping is efficient enough that most of the CR-scattering small-scale Alf\'en waves are thermalized and do not provide an efficient source of CR scattering any more. This leaves the CR transport parallel to the magnetic field uncoupled and CRs are free streaming. In such a state, the CR diffusion coefficient should approach infinity but the effective CR diffusion coefficient does not. Contrary,  effective CR diffusion becomes slower while going to the highest densities in our simulation.  

\section{Intrinsic vs.\ effective diffusion close to the disk}
\label{sec:intrinsic_vs_effectiv_near_disk}

\begin{figure*}
   \centering
   \resizebox{\hsize}{!} { \includegraphics[width=\textwidth]{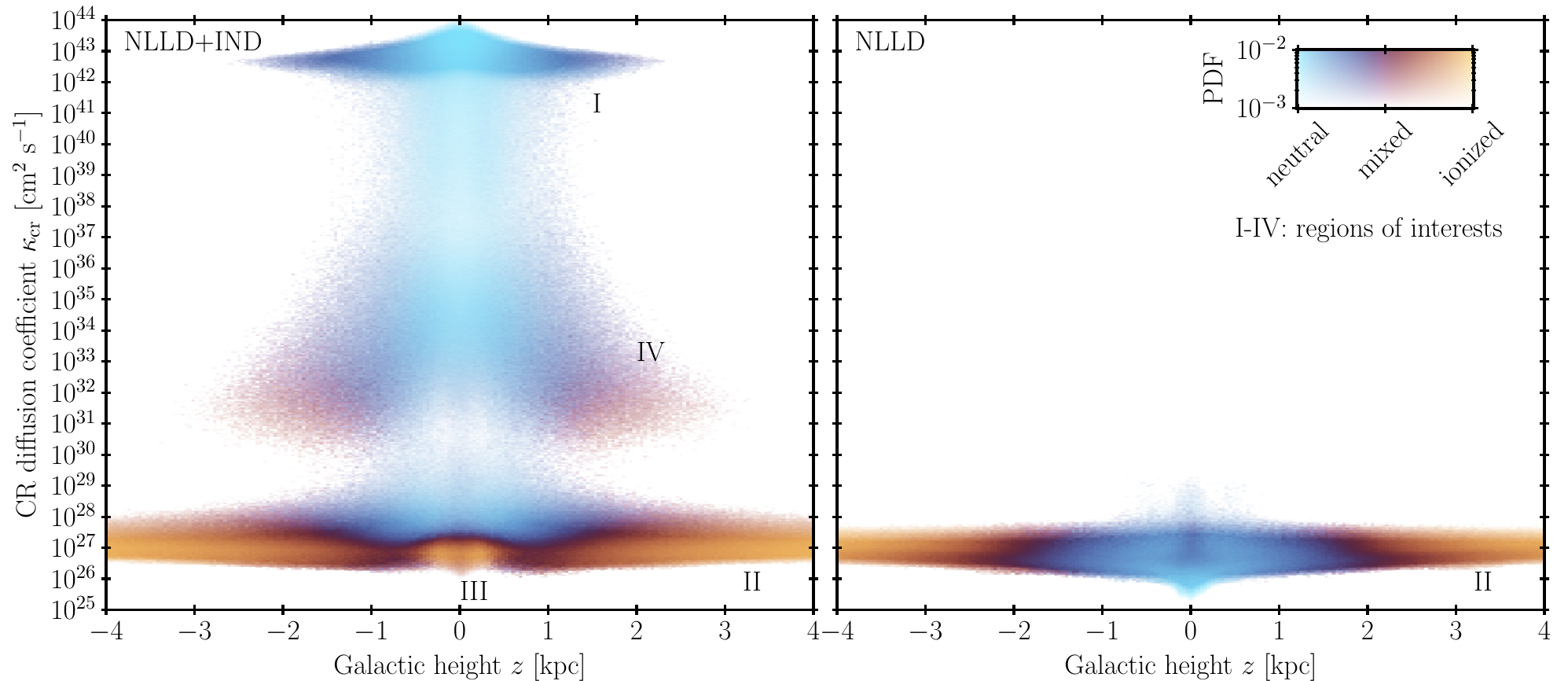} }
   \caption{Histogram highlighting the effect of ion-neutral damping on the intrinsic CR diffusion coefficient $\kappa_\mathrm{cr}$ as a function of galactic height (galactic midplane at $z=0$ can be found in the centre of each panel) for both the NLLD+IND (left panel) and the NLLD simulation (right panel). We colour each bin using the average ionization degree and determine the colour saturation in each bin using CR energy-weighted PDFs. We exclude Alfv\'en-wave dark regions by selecting only region where the CR streaming instability should be operating, where $f_\mathrm{cr} > 1.01 \varv_\mathrm{a,i} (\varepsilon_\mathrm{cr} + P_\mathrm{cr})$. Bins with yellow and blue colours are dominated by CRs in either predominantly ionized or neutral, while brownish colours indicate parameter space regions where CRs are found in both ionized and neutral regions. Light and white colours mean that this point in the parameter space is sparsely populated by CRs or does not contain CRs at all. Roman numerals mark special regions of interest which are discussed in the text.}
   \label{fig:multiphase-diffusion}
\end{figure*}

To directly see the impact of ion-neutral damping on the intrinsic CR diffusion coefficient $\kappa_\mathrm{cr}$, we plot this quantity in Fig.~\ref{fig:multiphase-diffusion} over the galactic height using a colour-coded CR-energy weighted PDF. While the saturation of the colours in the Fig.~\ref{fig:multiphase-diffusion} determines the value of the PDF, the hue of the colours reflects the CR-energy weighted ionization degree. We calculate this quantity based on the ionization state $f_i = n_i / n_\mathrm{tot}$, where $n_i$ is the ion number density and $n_\mathrm{tot}$ is the total number density, of each computation cell. For each pixel in the histogram, we calculate the CR energy in ionization weighted energies $E_\mathrm{ion} = \sum f_i E_\mathrm{cr}$ and $E_\mathrm{ion} = \sum (1 - f_i) E_\mathrm{cr}$ and calculate an overall ionization parameter
\begin{equation}
    C = \frac{E_\mathrm{ion} - E_\mathrm{neu}}{E_\mathrm{ion} + E_\mathrm{neu}},
\end{equation}
which is equal to -1 for neutral dominated regions and 1 for fully ionized regions. The value of this ionization parameter is used to interpolate between blue (neutral) and gold (ionized) colours. 

Alfv\'en-wave dark regions also have high intrinsic CR diffusion coefficients \citep{2023Thomas}. To avoid confusion between this phenomenon and ion-neutral damping, we only include CRs that are in principle able to excite the streaming instability (those CRs that are outside Alfv\'en-wave dark regions) by filtering for
$\vert \varv_\mathrm{cr} \vert > 1.01 \varv_a$. We include CRs that are within $R < 15$ kpc and are thus within the ISM of the galactic disk. 

To discuss this figure, we mark specific features in the PDF using Roman numerals:

\textbf{Region I}: These are predominantly neutral and have exceedingly high intrinsic CR diffusion coefficients. These are the gas parcels in the inner CGM and the ISM that are heavily affected by ion-neutral damping to the extent that all CR-scattering small-scale waves have been eradicated. This region can be found only in the NLLD+IND and not in the NLLD simulation because of our choice of included physical processes.

\textbf{Region II}: Gas parcels found in this parameter space region are ionized and can be found above the galactic disk. Here, non-linear Landau damping sets the diffusion coefficients, leading to low values of the intrinsic CR diffusion coefficient.

\textbf{Region III}: CRs in the ISM close to the galactic midplane. In the NLLD+IND simulation, the lowest intrinsic CR diffusion coefficient can be found in ionized material where no ion-neutral damping takes place. In the NLLD simulation, the ionization state of the gas has little impact on CR transport properties, such that low intrinsic CR diffusion coefficients can be found in predominantly neutral media. As non-linear Landau damping is a thermally induced damping process, it becomes less effective in cold media. This gives rise to the lowest intrinsic CR diffusion coefficients in highly neutral (and thereby cold) gas parcels close to $z\sim 0~$kpc.

\textbf{Region IV}: Because of numerical mixing at the boundary layers between physical regions where ion-neutral damping is active or not, some gas cells in our simulation can be found at intermediate values of the intrinsic CR diffusion coefficients. A similar behaviour was already noted in \citet{2023Thomas} but for the boundary layer at the rim of Alfv\'en-wave dark regions. We speculate that the presence of these data points in our simulations are purely numerical and are caused by insufficient numerical resolution, which leads to unresolved gradients in CR- and Alfv\'en wave-related quantities. 

\begin{figure}
   \centering
   \resizebox{\hsize}{!} { \includegraphics[width=\textwidth]{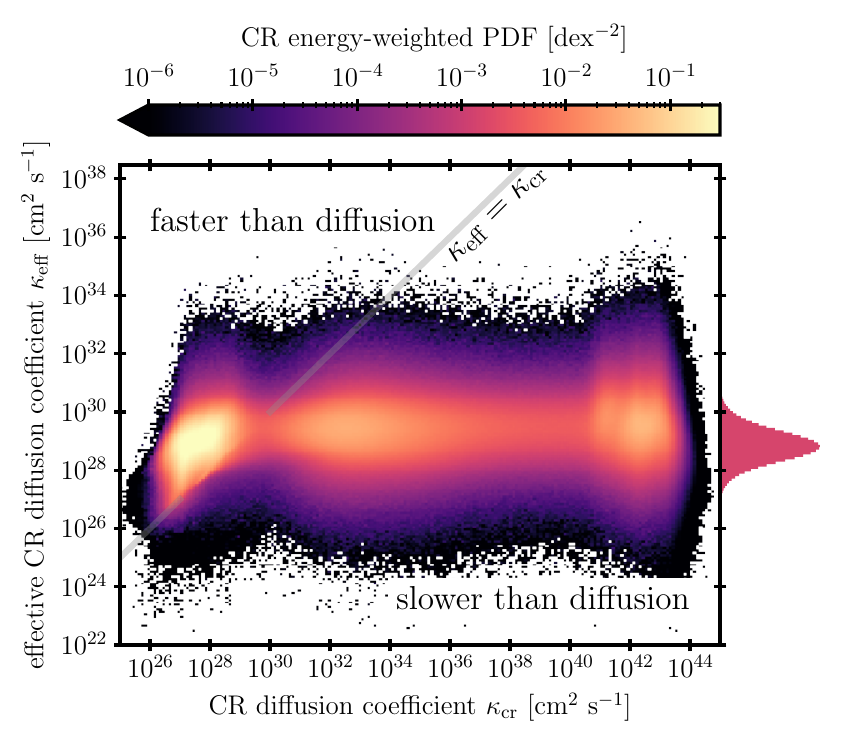} } \caption{Histogram displaying the joint distribution of the intrinsic CR diffusion coefficient $\kappa_\mathrm{cr}$ and the effective CR diffusion coefficient $\kappa_\mathrm{eff}$, defined in Eq.~\eqref{eq:kappa_eff}, weighted with the CR energy in the NLLD+IND simulation. The logarithm of the marginal distribution of the effective CR diffusion coefficient is shown on the right. The value of the effective CR diffusion coefficient appears to be independent of the intrinsic value. Because the effective CR diffusion coefficient is a measure of the speed of CR transport, we conclude that CR diffusion only has a little influence on CR transport and that the population of CRs with high intrinsic diffusion coefficients $\kappa_\mathrm{cr}$ is still describable by a low effective CR diffusion coefficient. }
   \label{fig:effective_kappa}
\end{figure}

The presence of very-high intrinsic CR diffusion coefficients at lower galactic heights in Fig.~\ref{fig:multiphase-diffusion}, has no counterpart in the vertical distribution of the effective CR diffusion coefficients in Fig.~\ref{fig:vertical_distribution}. We directly compare the two quantities in Fig.~\ref{fig:effective_kappa}, where we show the CR energy-weighted PDF and include data points from the last 100 Myr of evolution in the NLLD+IND simulation. This plot, again, contains all CRs irrespective of whether they are slower or faster than the Alfv\'en speed. We mark with a gray line where both CR diffusion coefficients coincide, $\kappa_\mathrm{cr} = \kappa_\mathrm{eff}$, and note that part of this line in Fig.~\ref{fig:effective_kappa} is omitted to reveal this specific feature in the actual simulation data. Close to this line, field-aligned CR transport is dominated by diffusion. The majority of CRs are effectively diffusing faster than predicted by the intrinsic CR diffusion coefficients; they can be found above the $\kappa_\mathrm{cr} = \kappa_\mathrm{eff}$ line. This is the case because, intrinsic CR diffusion has no notion of CR streaming which is additive to the CR diffusion process \citep{2020Thomas}. This transport effect increases the effective CR diffusion coefficient. The distribution also has a tail to the very high values of the intrinsic CR diffusion coefficients. CR energy that is found in this region of the parameter space is effectively transported slower than what would be predicted by a diffusion model that is based on the intrinsic CR diffusion coefficient. We see that these high values have hardly an impact on the actual ``effective'' CR diffusion coefficients, which share the same values as found for CRs having lower intrinsic CR diffusion coefficient values. In the regime of extremely high intrinsic CR diffusion coefficients, the CR diffusion approximation breaks down.

\section{Discussion and Outlook}
\label{sec:discussion}

The importance of ion-neutral damping in the context of CR transport has already been conjectured in the pioneering work of \citet{1969Kulsrud}. Our results confirm their conjecture, demonstrating that ion-neutral damping rapidly suppresses CR scattering by Alfv\'en waves in the ISM whenever a substantial population of both neutrals and ions is present. Our results are in line with the previous findings which focused on high resolution tall-box simulation \citep{2021Armillotta, 2022Armillotta, 2024Armillotta}. These \textsc{Tigress} simulations were post-processed using the two-moment CR transport model of \citet{2018Jiang} and similar reductions of the CR diffusion coefficient over many orders of magnitude inside the ion-neutral damping affected ISM. While their simulation focused on a kpc-sized patch of the ISM, our simulation covers the entire galactic disk and accurately captures the wind geometry. Nevertheless, theoretical expectations, high-resolution tall-box simulations, and global galaxy simulations all highlight the  major impact of ion-neutral damping on CR transport within the ISM.

An avenue left unexplored in this work is how CRs are transported in low metallicity environments. There, the thermochemistry is different compared to our work, where solar metallicity is effectively the floor metallicity, due to the altered impact of metal line cooling and dust mediated processes. The balance between the volume and mass fractions of the cold, warm, and hot phases is altered and depends on the effectiveness of CR-mediated heating \citep{2025Brugaletta}. Furthermore, wind driving above the galactic disk in our (solar-metallicity) model is driven by the CR gradient \citep{2025Thomas} while recent \textsc{Silcc} tall-box simulations \citep{2025BrugalettaII} show that CR and thermal pressure forces cancel one another and that the outflow is solely accelerated by magnetic fields in low metallicity regions. Both results show that CR-affected physics significantly depends on metallicity. Our present work shows that CR transport inside the ISM is heavily influenced by ion-neutral damping and thereby the abundance and balances of ions and neutrals. As this balance is altered in low-metallicity environments, an investigation of CR transport in such media is warranted.

Analysing a suite of \textsc{Fire-2} simulations, \citet{2021HopkinsIII} investigate the influence of CR-transport realizations on CR observables such as $\gamma$-ray luminosity and CR grammage by varying the applied physical models and used parameters. They conclude that observational constrains can be reproduced for transport models which achieve effective diffusion coefficients around $10^{29}$ to $10^{31}~\mathrm{cm}^2~\mathrm{s}^{-1}$ inside the ISM. Notably, one of their tested CR-transport variations employs the self-confinement picture, follows CR transport using our theory \citep{2019Thomas}, and accounts for ion neutral damping of Alfv\'en waves (next to other damping processes). This model is similar to the one presented here but ultimately fails to meet observational expectations. Interestingly, the average effective (isotropic) diffusion coefficient in their model is $2\times10^{27}\mathrm{cm}^2\mathrm{s}^{-1}$, and radial profiles show that for Milky Way-type galaxies, the effective diffusion coefficients start to surpass $10^{29}\mathrm{cm}^2\mathrm{s}^{-1}$ at galactocentric radii of $\sim$$100~\mathrm{kpc}$. Our present simulation can be characterized with effective diffusion coefficients of $\sim10^{29}~\mathrm{cm}^2~\mathrm{s}^{-1}$ throughout the whole galactic disk and the inner CGM (see Figs.~\ref{fig:effective_kappa} and \ref{fig:vertical_distribution}). The origin of this discrepancy, whether it is the missing cosmological context on our side or whether is stems from differences in the numerical modelling, is unknown. Furthermore, \citet{2024Ponnada} show that the ISM in  \textsc{Fire-2} simulations can be ``blown apart'' due to excessive overpressurization of self-confined CRs. We do not observe such an over-confinement of CRs in our simulations.

Our results notably rely on atomic collision rates for the momentum transfer between ion and neutrals in the ISM. While for some of the interactions parametrizations for the rates are based on quantum mechanical calculations, others can only be retrieved from the classical Langevin approximation. In particular, the \ion{H}{I}--\ion{C}{II} interaction is one of the important collisions setting the ion-neutral damping rate because of the vast availability of neutral hydrogen and singly ionized carbon in the diffuse ISM (\ion{C}{II} is abundantly produced by far UV radiation and can be found outside shock-ionized or \ion{H}{II}-regions). Currently, only the Langevin rate is available for this interaction, but a quantum-mechanical treatment is required. The collision is fundamentally quantum, involving excitation and charge-transfer processes that directly influence momentum transfer and, therefore, the ion–neutral damping rate.

\section{Summary \& Conclusions}
\label{sec:conclusion}

We present two idealized simulations of Milky Way-mass galaxies and investigate the impact of ion-neutral damping on CR transport inside the ISM and the inner CGM. To facilitate this, we run an isolated disk model using the same initial conditions for both simulations and run one simulation with active ion-neutral damping while deactivating it in the other. To model the ion-neutral damping process, we equip the \textsc{Crisp} galaxy formation framework with new rate coefficients that allow us to follow the damping provided by the most abundant primordial and metallic elements. This process occurs in the multiphase ISM, where ions and neutrals coexist, and it thermalizes most small-scale magnetic waves (assumed to be Alfv\'en waves) in our simulations. These waves normally scatter CRs and set the CR diffusion coefficient. We model CR transport using our two-moment hydrodynamical description \citep{2019Thomas} which follows not only the energy content of GeV CRs but also evolves the effective transport speed of CRs as an independent quantity. CRs in our simulations drive a galactic wind \citep{2025Thomas}, which is highly turbulent and enriched with CRs (see Fig.~\ref{fig:gallery_1}). Similar to our previous work, this galactic wind is filled with Alfv\'en-wave dark regions  where CRs damp all Alfv\'en waves which leaves them unscattered \citep{2023Thomas}. The resulting values for the intrinsic CR diffusion coefficient (a measure of how fast CRs diffuse in a Brownian-motion-like manner; see discussion in Sec.~\ref{sec:intrinsic_vs_effective}) start to diverge (see Fig.~\ref{fig:gallery_2}). 

In order to be able to characterize CR transport in Alfv\'en-wave dark regions and the ISM, we focus our investigation on the effective CR diffusion coefficient (see Eq.~\ref{eq:kappa_eff}). This quantity characterizes the CR transport speed along magnetic field lines and is expressed in units of an ordinary diffusion coefficient. Unlike the intrinsic CR diffusion coefficient, it does not exhibit spurious divergences in regions where CRs are not undergoing Brownian motion. The effective CR diffusion coefficient also has the useful property that, if CR transport is dominated by CR diffusion, the extrinsic and intrinsic values coincide. The effective diffusion coefficient is thus a true generalization of the idealized diffusion process and includes other transport modes such as streaming. 

Ion–neutral damping strongly affects CR transport in the ISM, as it eliminates nearly all CR-scattering Alfv\'en waves throughout most of its volume -- with the exception of the SN-driven hot (super-)bubbles where collisional ionization is strong enough to maintain the whole region ionized (see Fig.~\ref{fig:gallery_2} and \ref{fig:gallery_kappa_comparision}). This has profound consequences for CR transport. It increases the average CR transport in high-density patches of the ISM once the phase transitions from predominantly ionized to partially ionized and finally to fully neutral occur at densities $\rho \gtrsim 10^{-2}~m_\mathrm{p}~\mathrm{cm}^{-3}$. In the simulation affected by ion-neutral damping, CR transport speeds increase with higher density, whereas in the companion simulation without ion-neutral damping, transport slows at higher densities (see Fig.~\ref{fig:rho_vcr_correlation}). The effective diffusion coefficient exhibits the same trend across the simulations and shows enhancement within the ISM in the simulation that additionally accounts for ion-neutral damping. Overall, it declines with increasing density (see Fig.~\ref{fig:rho_kappa_eff_correlation}). We provide a fit for the median effective diffusion coefficient in Eqs.~\eqref{eq:kappa_eff_fit} and \eqref{eq:kappa_eff_fit_param}. Inclusion of ion-neutral damping has little impact on the CR-transport dynamics of low-density or high altitude environments.

The intrinsic CR diffusion coefficient diverges in neutral regions where Alfv\'en waves are effectively damped (see Fig.~\ref{fig:multiphase-diffusion}). In such regions, it is a poor measure for CR transport and shows no correlation with the effective CR diffusion coefficient (see Fig.~\ref{fig:effective_kappa}). The fast transport of CRs inside the ISM enables a shorter escape times of CRs from within the galactic plane into the inner CGM. This effect flattens the $P_\mathrm{cr} - \rho$ at ISM densities and lowers the median CR pressure therein (see Fig.~\ref{fig:rho_Pcr_correlation}). While ion-neutral damping heavily affects CR transport, it leaves the SFR almost unaltered (see Fig.~\ref{fig:star_formation_rate}). We close by emphasizing the importance of accurately simulating the multiphase ISM for correctly modelling CR transport and feedback in galaxies.

\begin{acknowledgements}
TT, CP, RL and MS acknowledge support by the European Research Council under ERC-AdG grant PICOGAL-101019746. The authors gratefully acknowledge the computing time granted by the Resource Allocation Board and provided on the supercomputer Emmy/Grete at NHR-Nord@Göttingen as part of the NHR infrastructure. The calculations for this research were conducted with computing resources under the project bbp00070.
This work was supported in part by the Perimeter Institute for Theoretical Physics.  Research at Perimeter Institute is supported by the Government of Canada through the Department of Innovation, Science and Economic Development Canada and by the Province of Ontario through the Ministry of Economic Development, Job Creation and Trade.
M.S. receives additional support through the Horizon AstroPhysics Initiative (HAPI), a joint venture of the University of Waterloo and Perimeter Institute for Theoretical Physics.

\end{acknowledgements}

% WARNING
%-------------------------------------------------------------------
% Please note that we have included the references to the file aa.dem in
% order to compile it, but we ask you to:
%
% - use BibTeX with the regular commands:
%   \bibliographystyle{aa} % style aa.bst
%   \bibliography{Yourfile} % your references Yourfile.bib
%
% - join the .bib files when you upload your source files
%-------------------------------------------------------------------

\bibliographystyle{aa}
\bibliography{main}

\begin{appendix} 

\end{appendix} 

\end{document}